# Investigating the Electronic Structure of Prospective Water-splitting Oxide BaCe$_{0.25}$Mn$_{0.75}$O$_{3-\delta}$ Before and After Thermal Reduction

Subhayan Roychoudhury§, Sarah Shulda¶, Anuj Goyal¶, Robert Bell¶, Sami Sainio‡, Nicholas Strange‡, James Eujin Park#, Eric N. Coker#, Stephan Lany¶, David Ginley¶, and David Prendergast§

§ The Molecular Foundry, Lawrence Berkeley National Laboratory, Berkeley CA 94720, USA
¶National Renewable Energy Laboratory, Golden, Colorado 80401, USA
‡SLAC National Accelerator Laboratory, Menlo Park, CA 94025
#Sandia National Laboratories, Albuquerque, New Mexico 87185, USA

# Abstract

BaCe$_{0.25}$Mn$_{0.75}$O$_{3-\delta}$ (BCM), a non-stoichiometric oxide closely resembling a perovskite crystal structure, has recently emerged as a prospective contender for application in renewable energy harvesting by solar thermochemical hydrogen generation. Using solar energy, oxygen-vacancies can be created in BCM and the reduced crystal so obtained can, in turn, produce H$_2$ by stripping oxygen from H$_2$O. Therefore, a first step toward understanding the working mechanism and optimizing the performance of BCM, is a thorough and comparative analysis of the electronic structure of the pristine and the reduced material. In this paper, we probe the electronic structure of BCM using the combined effort of first-principles calculations and experimental O $K$-edge x-ray absorption spectroscopy (XAS). The computed projected density-of-states (PDOS) and orbital-plots are used to propose a simplified model for orbital-mixing between the oxygen and the ligand atoms. With the help of state-of-the-art simulations, we are able to find the origins of the XAS peaks and to categorize them on the basis of contribution from Ce and Mn. For the reduced crystal, the calculations show that, as a consequence of dielectric screening, the change in electron-density resulting from the reduction is strongly localized around the oxygen vacancy. Our experimental studies reveal a marked lowering of the first O $K$-edge peak in the reduced crystal which is shown to result from a diminished O-$2p$ contribution to the frontier unoccupied orbitals, in accordance with the tight-binding scheme. Our study paves the way for investigation of the working-mechanism of BCM and for computational and experimental efforts aimed at design and discovery of efficient water-splitting oxides.

## Introduction:

The quest for developing advanced mechanisms capable of harvesting energy in a renewable, sustainable and environment-friendly manner is one of the grand challenges of modern technological research. One such example is the production of solar fuel by splitting $H_2O$ or $CO_2$ through a reversible two-step cycle involving reduction and re-oxidation of metal oxides. In particular, in one of its most promising approaches, the process of Solar Thermochemical Hydrogen generation (STCH)[1–3] entails, as a first step, the removal of oxygen atoms from a metal oxide at a high temperature generated using solar irradiation under inert atmosphere. This is then followed by a re-oxidation of the oxide by exposing it to $H_2O$, resulting in the production of hydrogen. For efficient water-splitting, the oxide must be structurally stable to withstand a large number of redox cycles. Additionally, it must be reducible at relatively low temperature and undergo re-oxidation in spite of the presence of $H_2$ in the steam feed. The efficacy of an oxide in the STCH operation is governed by its chemical properties, which, in turn are primarily dictated by its electronic structure. Being able to decipher and manipulate the electronic structure of prospective water-splitting oxides with dedicated theoretical and experimental investigation is, therefore, of paramount importance in gaining a deeper understanding of their working-mechanism as well as in the discovery, design and performance-optimization of such materials.

Owing to their stability against decomposition and phase-change under the extreme environment of a thermochemical reactor, non-stoichiometric oxides have recently received substantial attention in this regard[4]. In particular, non-stoichiometric ceria ($CeO_2$) is generally recognized as the frontrunner among the prospective candidates[5,6] since, besides possessing other beneficial attributes, it is capable of stripping oxygen from $H_2O$ even under an extremely unfavorable $H_2O:H_2$ ratio[4] . However, contrary to the ease of the re-oxidation step, the reduction step for ceria poses a major disadvantage since it requires a temperature above $1500^\circ C$[7,8], which necessitates non-standard materials and becomes prohibitively expensive under standard industrial conditions.

As a consequence, the last few years have witnessed the search for alternative oxides that can undergo reduction at a lower temperature, with significant attention being devoted toward perovskite-based structures[9–15]. Recently, the non-stoichiometric perovskite compound $BaCe_{0.25}Mn_{0.75}O_{3-\delta}$ (BCM) has been proposed as a contender for the STCH application[16]. In addition to high chemical stability and tunability of the point-defect thermodynamics, BCM is shown to offer nearly a three-fold improvement over ceria in hydrogen production at a reduction temperature below $1400^\circ C$. Additionally, like ceria, BCM is also capable of operating under unfavorable $H_2O:H_2$ ratio. Therefore, as a first step toward gaining a detailed understanding of the factors responsible for its performance, a thorough exploration of the electronic structure of BCM is necessary. In particular, it is imperative that such a study addresses the change in the electronic structure induced by the reversible removal of oxygen atom from the compound.

On the experimental front, investigation of the electronic structure of materials is performed most commonly using core-level spectroscopy. For example, x-ray absorption spectroscopy (XAS) produces an element-specific spectrum of the absorption intensity plotted as a function of photon-energy. However, in order to decipher the details of electronic structure from such a plot, the experimental efforts must often be complemented with theoretical studies. The aim of such theoretical calculations is then to be able to gain insight into the origin of the spectral features by assigning electronic processes and orbitals to them. For several decades, *ab initio* computation of electronic properties has been performed most commonly with the help of density-functional theory (DFT), thanks to its convenient balance between accuracy and expense. However, in addition to such first-principles simulations, simplified quantum-mechanical models can be extremely helpful in elucidating the electronic properties of most materials.

With these considerations, in this paper we present a detailed combined theoretical and experimental investigation of the electronic structure of BCM. The information obtained with first-principles DFT calculations are presented mostly in the form of partial density-of-states (PDOS) and real-space representation of single-electron energy eigenstates, referred to as electronic orbitals hereafter. The computational studies are then complemented with crystal-field theory (CFT)-based models of orbital mixing which corroborate the first-principles results qualitatively. Experimental O $K$-edge XAS is performed to probe the electronic excitations before and after oxygen-removal from the compound. The absorption spectra so obtained are compared with their simulated counterparts obtained using an accurate many-body method. In particular, prominent changes in the experimental spectrum upon oxygen removal are replicated by the computational results which, in turn, are shown to follow from simplified models.

## Computational Formalism:

Experimental and computational details, along with the crystal structure, are included in the Supplementary Materials. However, since a major portion of the paper deals with the analysis of x-ray absorption spectra, in order to facilitate the discussion, here we provide a brief outline of the computational framework employed in the spectroscopic simulations. The O $K$-edge spectra have been simulated using the Many-Body X-ray Absorption Spectroscopy (MBXAS)[17,18] formalism which, for an electronic excitation from the initial ground state (GS) $|\Psi_{GS}\rangle$ to some final core-excited state $|\Psi_f\rangle$, evaluates the transition amplitude $A_{GS \to f}$ as

$$A_{GS \to f} = \langle \Psi_f | \hat{O} | \Psi_{GS} \rangle,$$

where $\hat{O}$ is the many-body transition operator, which, in these calculations, is approximated by the dipole term. $|\Psi_{GS}\rangle$ is approximated by a Slater determinant (SD) composed of the occupied Kohn-Sham (KS) orbitals of the DFT GS. The SD for $|\Psi_f\rangle$, on the other hand, is constructed by populating KS orbitals of a different self-consistent field – that of a core-ionized state with an empty core orbital on a specific atom, which we refer to as the full core-hole[1] (FCH) KS system. A specific final state, $|\Psi_f\rangle$, refers to a SD comprising the valence occupied orbitals of the FCH system together with the specific (previously unoccupied) orbital indexed by $f$ (within the so-called f$^{(1)}$ approximation).

---

[1] Lowest energy state of the system corresponding to core-ionization of the relevant atom.

# Results and Discussions:

## (1)    Structure and Configuration

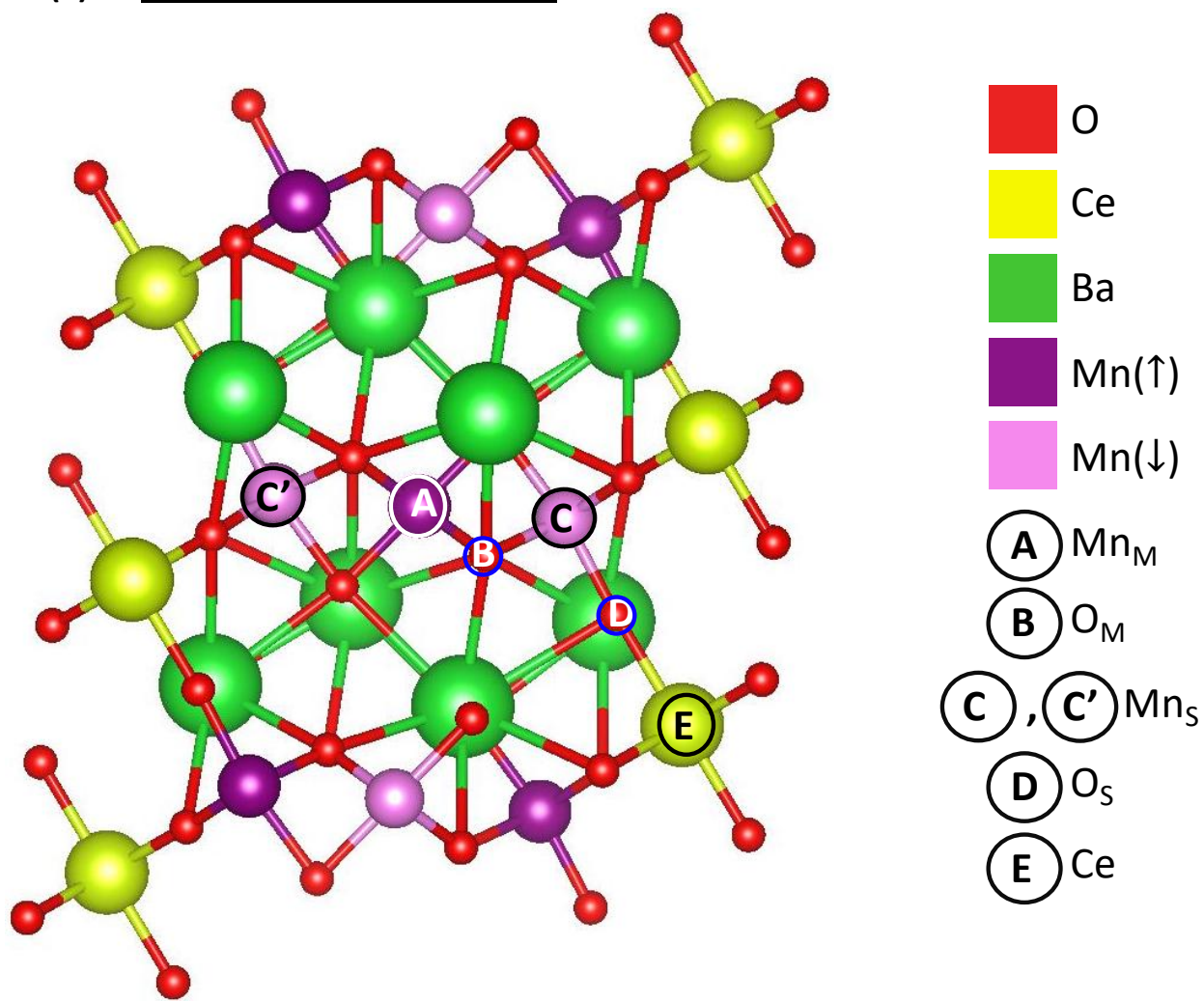

*Fig. 1 The structure of a pristine 12R-BCM crystal. The atoms labeled A, B, C (also C'), D and E are representatives of a $Mn_M$, $O_M$, $Mn_S$, $O_S$ and Ce atom, respectively. Atoms C', A and C collectively constitute a representative triple of Mn atoms, as mentioned in the text. Each triple of Mn atom contains two $Mn_S$ atoms of identical magnetic moment (C and C', for this particular triple) and an $Mn_M$ atom which has a magnetic moment in the opposite direction (A, for this particular triple).*

As shown in Fig. 1, 12R-BCM has a rhombohedral crystal structure[2] (R-3m, #166), with each of Ba, Mn, and O occupying the two unique Wyckoff sites and Ce a single Wyckoff site. Within the standard perovskite notation (not to be confused with the notation used in Fig. 1), the Ba atoms occupy the 'A' sites while Mn and Ce share the 'B' sites. Both Mn and Ce atoms exhibit six-fold coordination to neighboring O atoms. O atoms are present in two inequivalent coordinations each with two-fold coordination: $O_M$ refers to a ``Middle'' O atom, which is coordinated by two Mn atoms while $O_S$ refers to a ``Side'' O atom which is coordinated by a single Mn and a single Ce atom. Similarly, the Mn atoms exhibit two inequivalent coordinations in the lattice: those with a Ce atom as a second nearest-neighbor are hereafter labeled $Mn_S$, while those with two other Mn atoms as second nearest neighbors are labeled $Mn_M$. The Mn-O coordination environment is not perfectly octahedral because of the unequal Mn-O angles and bond lengths: 1.89 Å for $Mn_S$-$O_S$ and 1.99 Å (1.93 Å) for $Mn_S$-$O_M$ ($Mn_M$-$O_M$). Ce-O bonds are arranged in a perfect octahedral symmetry with a Ce-O bond length (2.23 Å) larger than the Mn-O bond length. Notably, the difference in the Mn-$O_S$ and the Ce-$O_S$ bond lengths (2.23 Å -1.89 Å , i.e., 0.34 Å) is exactly equal to the difference in the ionic radius of $Ce^{4+}$ (1.01 Å)[3] and $Mn^{4+}$ (0.67 Å)[4].

The Mn atoms, which are in the $Mn^{4+}$ oxidation state, have strong net magnetic moments due to the $3d^3$ valence occupancy and are arranged in triples (groups of three neighbors). In our DFT calculations, we simulate the lowest energy anti-ferromagnetic spin configuration of Mn atoms in the rhombohedral structure such that, in terms of the magnetic moment, the Mn atoms follow an alternating arrangement. Within every triple of Mn atoms, if the $Mn_M$ site has a net up (down) magnetic moment, then both of the $Mn_S$ sites will have the opposite down (up) magnetic moment, resulting, for each triple, in a net magnetic moment with the direction dictated by the 2:1 majority of $Mn_S$ constituents. In the particular anti-ferromagnetic arrangement in our chosen supercells, all triples within a given layer are spin-polarized along the same direction while triples belonging to neighboring layers have opposite spin-polarizations. The calculated ground-state total energy of such a polarization-arrangement is seen to be lower than that of others: e.g., the ferromagnetic arrangement where all Mn atoms are polarized along the same direction or the anti-ferromagnetic one in which neighboring triples differ in their net spin-polarization but within a given triple, all Mn atoms are polarized along the same direction.

## (2)    Electronic structure of the pristine crystal

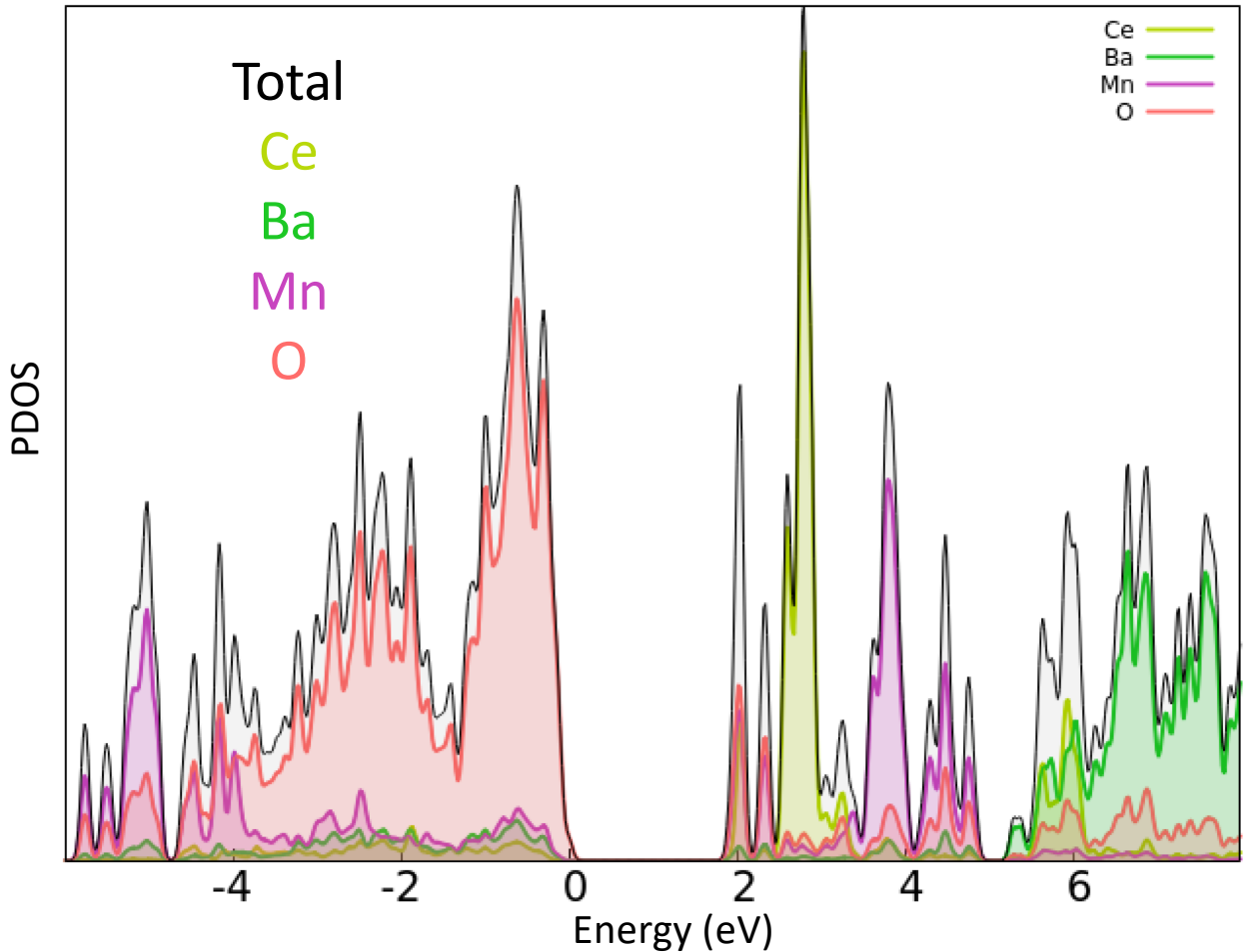

*Fig. 2 The total and partial DOS of the pristine BCM crystal. The energy of the VBM is set at zero.*

Fig. 2 shows the density of states (DOS) of the pristine BCM crystal, along with the projected DOS (PDOS) contributions from the different species. The small contribution of Ce-4$f$ in the occupied PDOS indicates that these orbitals do not partake significantly in mixing with O-2$p$ orbitals. The Ce-4$f$ states are mostly unoccupied, with significant peaks appearing above the Fermi level, and highly localized in energy; confined within 1.5 eV on the energy axis with peak value around 0.75 eV above the CBM. In fact, the PDOS at the LUMO energy shows comparable contributions from O-*p*, Mn-3*d* and Ce-4*f* orbitals. The energy of the Ce-4*f* band is relatively similar across the two spin channels because Ce in the $Ce^{4+}$ charge state has an empty valence 4*f* orbital.

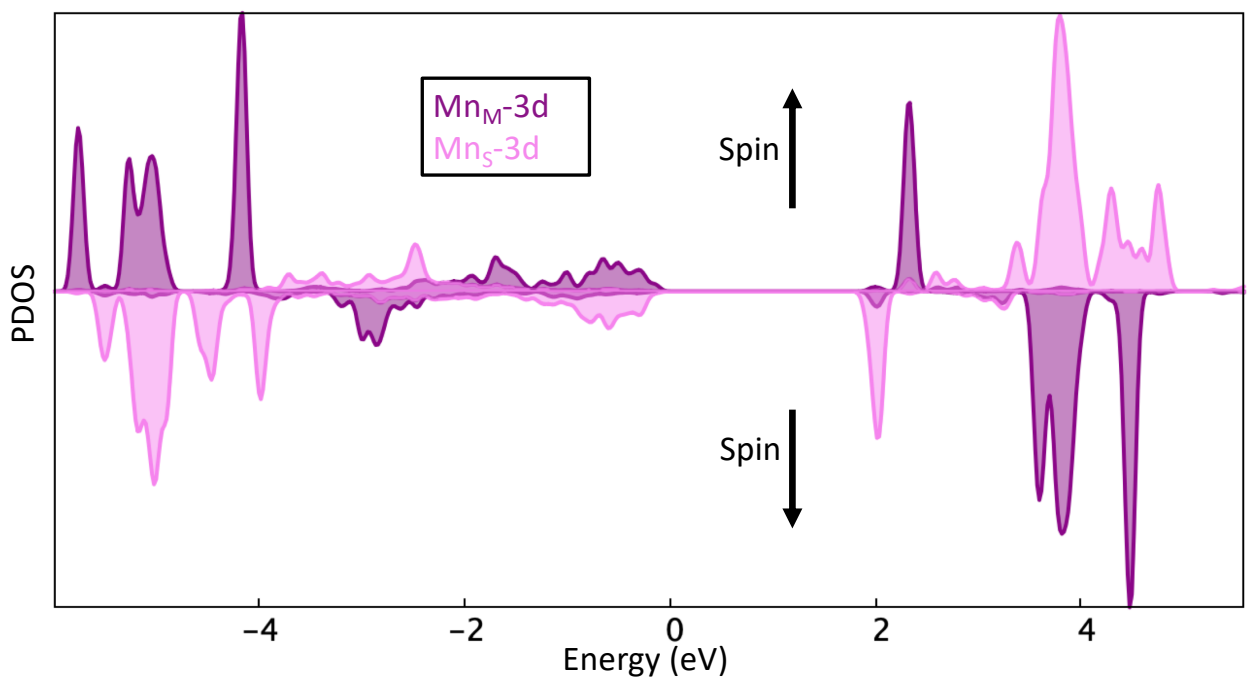

*Fig. 3 PDOS plot of an $Mn_M$ and an $Mn_S$ atom belonging to the same triple. In this representative triple, the $Mn_M$ and $Mn_S$ atoms have a net up and down spin-polarization, respectively [see, for example, the triple consisting of the atoms C', A and C in Fig. 1]. On the energy axis, 0 corresponds to the VBM.*

The magnitude of Lowdin polarization is seen to be lower (±3.13) for $Mn_M$ than it is for $Mn_S$ (±3.19), as reported in Tab. 1. This can be associated with the exchange interaction of $Mn_M$ with electrons from the second nearest-neighbors, both of which are $Mn_S$ atoms with magnetic moments in the direction opposite to that of the $Mn_M$ atom. Note from Fig. 3, which shows the 3-*d* PDOS of a $Mn_M$ and a $Mn_S$ atom (which are constituents of the same triple and therefore, are spin-polarized along opposite directions) that, in their respective majority spin channels, the CBM in the PDOS of $Mn_M$ is at a higher energy compared to that of $Mn_S$. This trend, however, is reversed for the lowest unoccupied peaks of their respective minority channels. Noting that a larger polarization typically corresponds to a larger exchange-splitting, which, in turn, would result, in accordance with the tight-binding formalism, in a larger energy-difference between the unoccupied levels of the two spin-channels, we can attribute the aforementioned property of the PDOS to the relatively higher (lower) local magnetic moment of $Mn_M$ ($Mn_S$).

We can begin to understand the electronic structure by making some initial simplifications. If we assumed a local octahedral coordination, the Mn 3$d$-orbitals for each spin channel would split into $e_g$ and $t_{2g}$ orbitals. The former would occupy a higher energy subspace as a consequence of increased electronic repulsion resulting from the alignment along the octahedral Mn-O bonding-axes, while the latter would be oriented between the bonding-axes and consequently, would have lower energy.

(i)   In our attempt at deciphering the electronic structure, as a first approximation, we can neglect all orbital-mixing between the Mn and the O atoms. Since the Mn atoms have a 3$d^3$ configuration, under such a hypothetical, fully-ionic $Mn^{4+}$ scenario, then, in the majority spin channel, we would expect each $t_{2g}$ orbital to be occupied with parallel spins (according to Hund's rule) with the $e_g$ orbitals remaining unoccupied. In the minority spin channel, the Mn 3$d$ orbitals would have higher energy due to exchange splitting and therefore all the minority $d$-orbitals would be completely empty, giving rise to a net absolute atomic magnetic moment of 3 $\mu_B$ for each Mn atom.

(ii)   Introducing a further layer of complexity, in accordance with the concepts of ligand-field theory within the solid state[19], we can take into account the mixing between the Mn-$e_g$ and the O-2$p$ orbitals, while retaining the non-bonding nature of the Mn $t_{2g}$ orbitals since the latter are not oriented along the bonding axes. The mixed (Mn-$e_g$ + O-2$p$) orbitals would then split into occupied bonding and unoccupied anti-bonding orbitals. Thus, in the majority spin-channel, the occupied orbitals would exhibit either a bonding (Mn-$e_g$ + O-2$p$) nature or a non-bonding Mn-$t_{2g}$ nature, with the VBM consisting predominantly of the former. The unoccupied orbitals can be expected to show (Mn-$e_g$ + O-2$p$) nature. In the minority spin channel, all occupied orbitals would show exclusively (Mn-$e_g$ + O-2$p$) bonding character while in the unoccupied subspace, there would be anti-bonding (Mn-$e_g$ + O-2$p$) and non-bonding Mn-$t_{2g}$ orbitals.

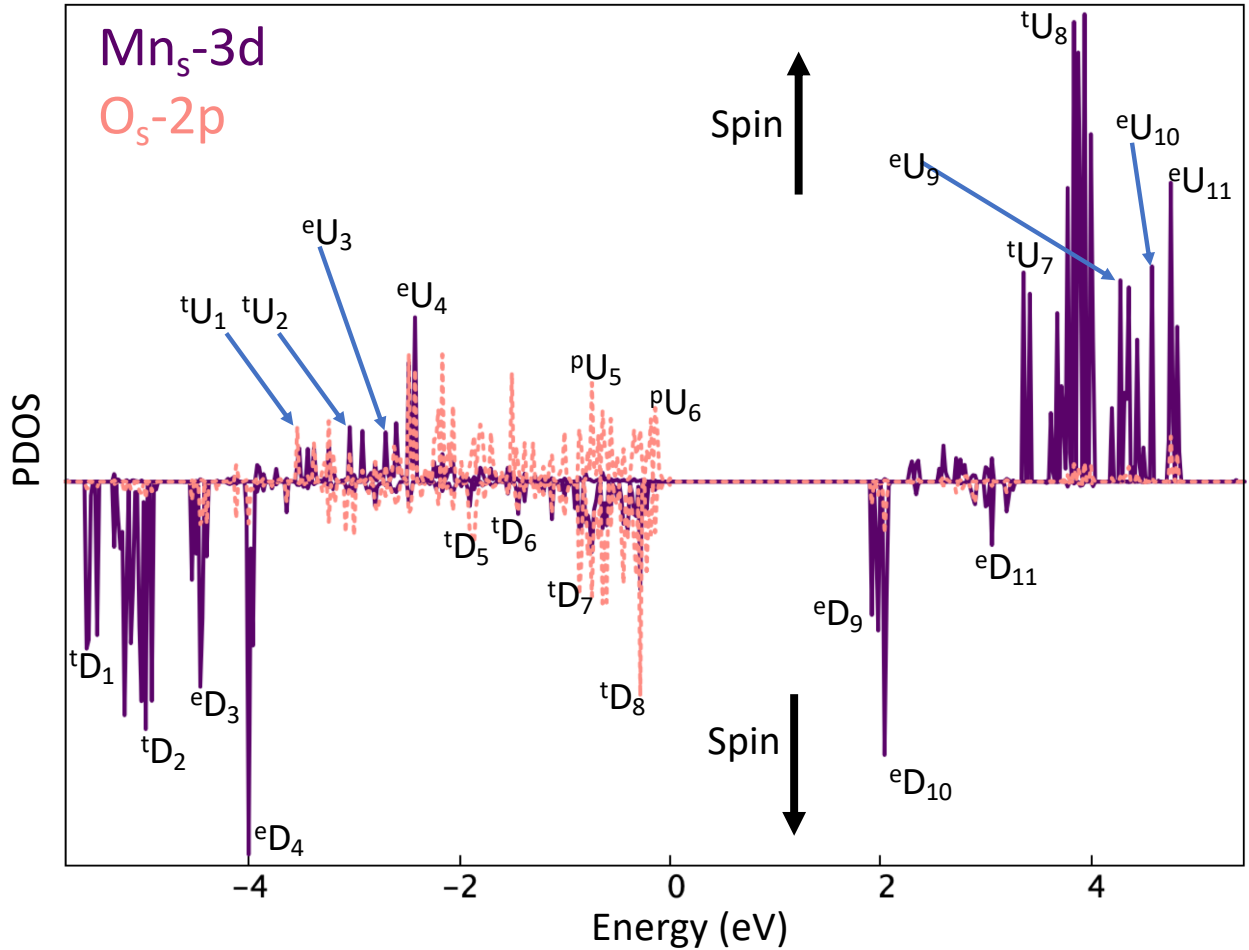

*Fig. 4 The PDOS of an $Mn_S$ and a neighboring $O_S$ atom, showing their 3d and 2p contributions, respectively. As in Fig. 3, the $Mn_S$ atom has a majority of down spin electrons. Certain peaks in the up (U) and down (D) spin channels of the PDOS are labeled for further analysi of the corresponding orbitalss, as shown with isovalue plots in Fig. 5 and Fig. 6. In the labels, the superscript t and e, refers to an orbital showing appreciable mixing of O-2p with Mn- $t_{2g}$ and Mn- $e_g$ , respectively, while the superscript p refers to an orbital having predominantly O-2p character.*

In order to test the validity of two orbital mixing approximation, we analyze the 2$p$ orbital PDOS of an $O_S$ atom and the 3$d$ orbital PDOS of the adjacent $Mn_S$ atom in

Fig. 4. Note that, in contrast with $O_M$, $O_S$ is coordinated with atoms of two separate species: an $Mn_S$ and a Ce, the latter of which shows much weaker mixing with O $p$-orbitals in comparison with the $Mn_s$ (as evident from the low contribution of Ce $4f$ in the occupied PDOS shown in Fig. 2), making the PDOS analysis of $O_S$-$2p$ comparatively simpler. A low numerical broadening (2.86 meV) has been used to facilitate the assignment of electronic orbitals to the PDOS features in Fig. 4. Fig. 5 and Fig. 6, present the isovalue plots of the KS orbitals contributing to various salient peaks of the PDOS for the up (U) and the down (D) spin channels, respectively. Notice, for example, that, in contrast with the aforementioned prediction of ligand-field theory, in the majority, i.e., down spin-channel, the PDOS around the VBM is dominated by orbitals exhibiting mixed ($Mn_S$-$t_{2g}$ + $O_S$-$2p$) character [e.g. $^tD_7$, $^tD_8$] while orbitals at much lower energy (-4 eV) show distinct ($Mn_S$ -$e_g$ + $O_S$-$2p$) mixing [e.g. $^eD_3$, $^eD_4$]. Additionally, in the minority, i.e., up spin-channel, the occupied orbitals $^tU_1$, $^tU_2$ show noticeable ($Mn_S$-$t_{2g}$ + $O_S$-$2p$) character. In order to explain such anomalies,

(iii)     we propose in Fig. 7, an alternative mixing scheme consistent with the orbital-characteristics observed in Fig. 5 and Fig. 6 which accounts for the possibility of mixing between the $Mn_S$ -$t_{2g}$ and the $O_S$-$2p$ orbitals, which is possible, especially due to the fact that the Mn-O coordination environment is not perfectly octahedral. This additional mixing leads to splitting of the $Mn_S$ -$t_{2g}$ manifold into: a lower energy band of what would be called non-bonding orbitals under the mixing scheme (ii) outlined above, but which here exhibit some mixing with $O_S$-$2p$; and a higher energy band, the anti-bonding counterpart, which, depending on the spin channel, may be occupied (down spin, majority) or not (up spin, minority).

Fig. 5 also shows significant $O_S$-$2p$ (non-bonding) character near the up-spin VBM [$^pU_5$ and $^pU_6$]. Additionally, from Fig. 6, we can notice substantial presence of Ce-$4f$ in the KS orbitals $^eD_{10}$ and $^eD_{11}$, which is consistent with the PDOS plot shown in Fig. 2. Overall, the schematic picture in Fig. 7 provides a consistent view of the ultimate electronic structure observed in our simulations.

In summary, we have seen that the Ba orbitals have very little presence in the frontier PDOS. The Ce-$4f$ orbitals, which show weak mixing with O-$2p$, are predominantly localized near the conduction band edge. The Mn-$3d$ orbitals, on the other hand, mix substantially with O-2$p$, generating bonding and anti-bonding levels across a wide energy range, as shown schematically in Fig. 7.

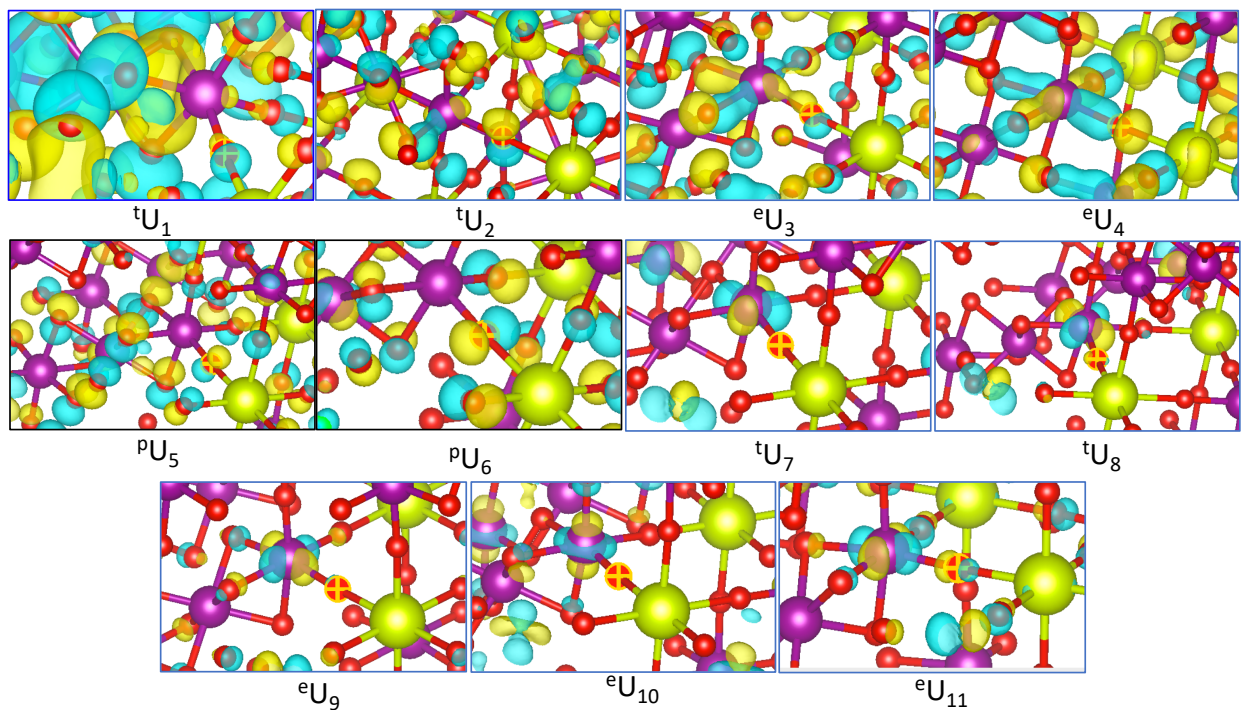

*Fig. 5 Isovalue plots of up spin KS orbitals having significant contributions to the peaks labeled in Fig. 4. The relevant $O_S$ atom is marked with a "+" symbol.*

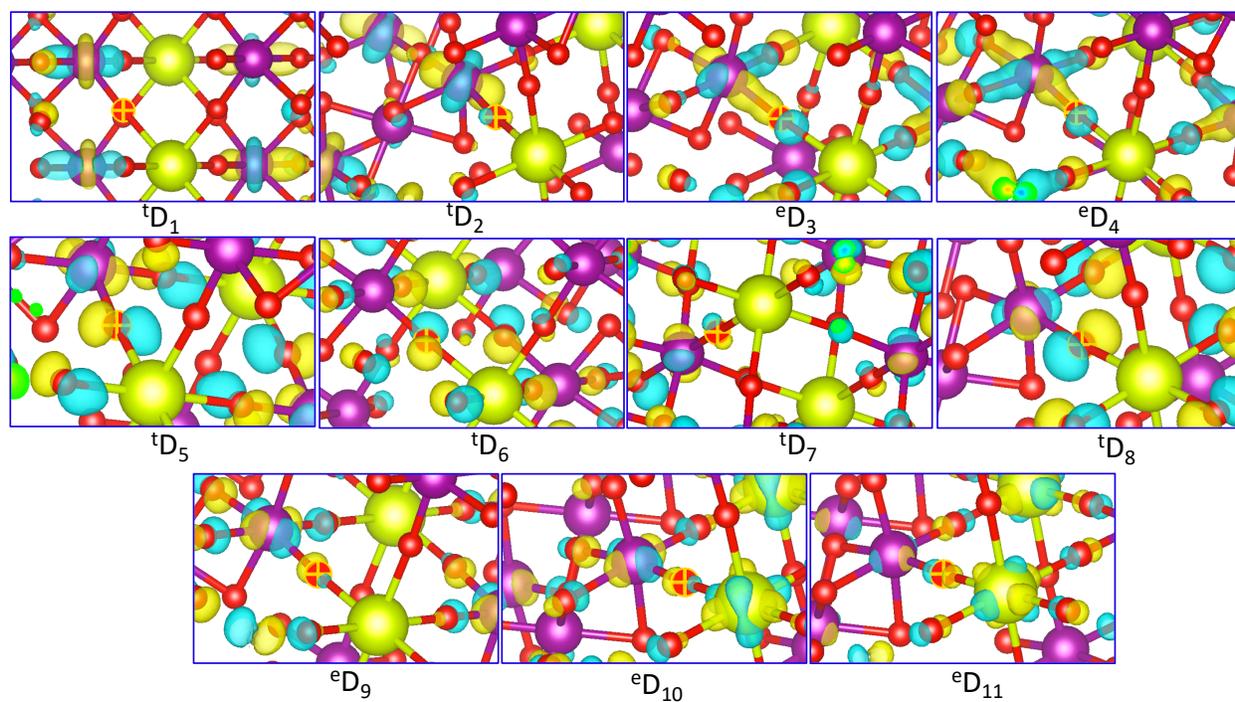

*Fig. 6 Isovalue plots of down spin KS orbitals having significant contributions to the peaks labeled in Fig. 4. The relevant $O_S$ atom is marked with a "+" symbol.*

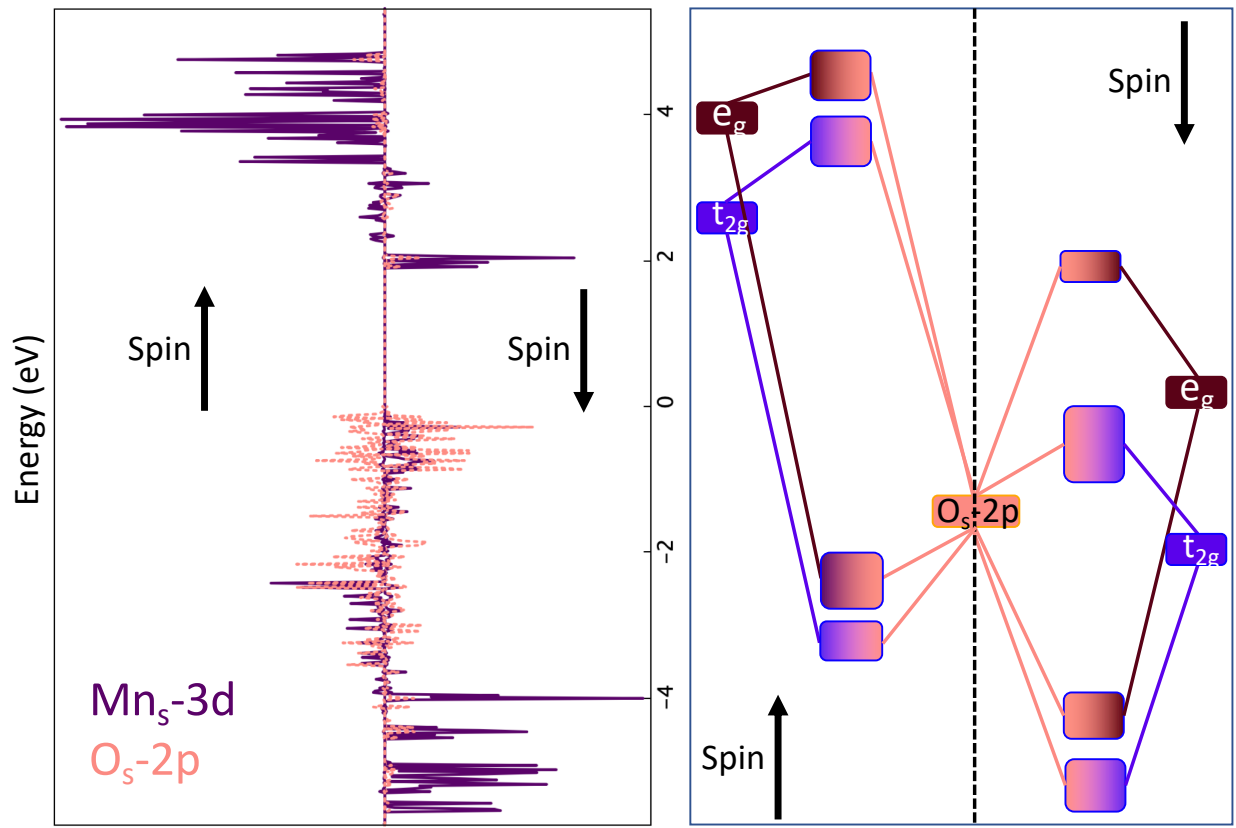

*Fig. 7 PDOS of $Mn_S$-3d and $O_S$-2p, along with a schematic detailing the bonding scheme in accordance with the LCAO formalism. Note that the down (up) spin channel corresponds to majority (minority) spins.*

## (3) O K-edge XAS of the Pristine Crystal

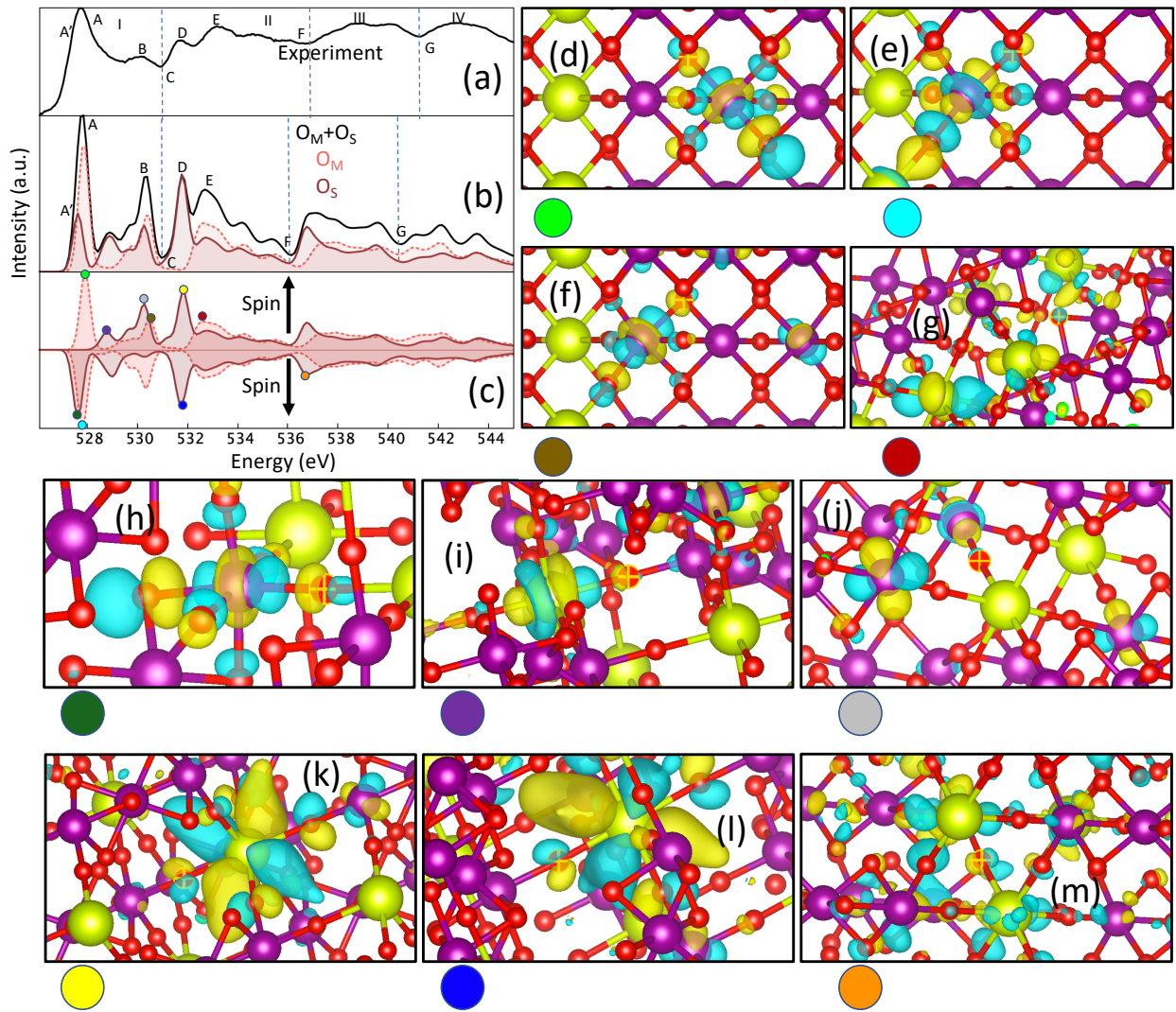

Fig. 8 Top left panel shows the O K-edge XAS of the pristine BCM crystal. Panel (a) shows the experimental spectrum while the panel (b) shows the resultant simulated spectrum, along with the individual contributions from $O_S$ and $O_M$. In order to facilitate comparison, the spectra have been divided into four regions along the energy axis and the salient crests and troughs have been labeled. Panel (c) shows, for both of the oxygen atoms, the contribution of the up and the down spin channels to the net spectrum. The rest of the panels show isovalue plots of orbitals of the promoted electron in the excited-states resulting from core-excitation of an $O_M$ (panels (d)-(g)) or an $O_S$ atom (panels (h)-(m)). The core-excited atom is marked by a "+" symbol. The XAS peak marked by circle of a given color in panel (c) has One of the most efficacious avenues for experimental investigation of electronic structure is the element-specific X-ray spectroscopy. In the following, we present a detailed experimental and theoretical account of the O $K$-edge X-ray absorption spectra (XAS) associated with exciting a $1s$ core electron of an O atom. Since the BCM unit cell contains an equal number of $O_M$ and $O_S$ atoms, the final spectrum will be the resultant of contributions from both of these atoms. Notably, the final state created by the process of X-ray absorption differs from the GS in that it has a missing 1s electron (a hole) localized on a specific oxygen atom. In addition, the core-hole creation results in a significant modification of electron-density in the vicinity of the excited atom $O^{(x)}$, resulting, in accordance with the tight-binding framework, in an enhanced contribution to the occupied PDOS and a downward shift in the energy of those orbitals, due to strong, attractive Coulomb interactions with the core hole. Concomitantly, the contributions of the excited atom to the unoccupied PDOS, due to orbitals of mixed character with neighboring atomic sites may be reduced. A comparison of the PDOS of individual atoms corresponding to the ground and the FCH state is presented in Fig.S1. The FCH state is defined above and is the foundational self-consistent field from which the core-excited Kohn-Sham orbitals are obtained.

The experimental O $K$-edge spectrum of BCM is presented in Fig. 8(a). As mentioned in the introduction, the goal then, is to extract meaningful electronic-structure information from the spectrum by associating electronic excitations to the salient spectral features. As a first step to this end, in Fig. 8(b) we show the simulated resultant O $K$-edge spectrum, as well as the individual contributions from $O_M$ and $O_S$. As expected, $O_M$, which is coordinated by two Mn atoms with opposite magnetic moments of comparable magnitude, produces relatively symmetric contributions to the absorption spectrum for the two spin channels (Fig. 8(c)). On the other hand, the spectral contributions of $O_S$, which is coordinated by a Mn and a Ce atom having significant and negligible magnetic moments respectively, differ considerably between the spin channels. This asymmetry leads to a lower intensity of the first

XAS peak in the resultant spectrum of $O_S$. From Table 2, $O_S$ can be seen to have a slightly higher Lowdin magnetic moment, indicating a stronger spin-polarization, than $O_M$. The exchange interaction arising from this higher spin-polarization lowers the conduction orbitals of the majority spin channel associated with $O_S$, as evident from the slightly lower energy of the first XAS peak of $O_S$ (compared to $O_M$) in the down spin channel. It is important to note that the appearance of the simulated spectrum is heavily dependent on the broadening used for generating it. A spectral peak which is well-separated in energy from the rest of the features will typically experience a substantial lowering of intensity with increment in broadening. In Fig. 8 (b)-(c), a uniform broadening of 0.2 eV is used, while the spectra simulated with higher broadening energies are shown in Fig. S2 in the Supplementary Materials. In particular, Fig. S2 reveals that the relative intensities of the net simulated spectrum depend on the broadening. For higher broadening values, the relative intensity of the intense peak A diminishes in comparison with the others. Additionally, the features at 528.9 eV and 529.7 eV, which are present between A and B in in the simulated spectrum (black line in Fig. 8(b)), but not visible in the experimental counterpart (Fig. 8(a)), vanish for higher values of broadening.

| Peak | O atom | Ligand atom | Note |
|------|--------|-------------|------|
| A' | $O_S$ | $Mn_S$ | ($O_S$-2p+$Mn_S$-3d) orbital in majority spin-channel |
| A | $O_M$ | Mn | ($O_M$-2p+Mn-3d) orbital in majority spin-channel. Both up and down orbitals contribute, since $O_M$ has two neighboring Mn atoms with opposite spin-polarization. |
| B | O | Mn | ($O_S$-2p+$Mn_S$-3d) orbital in the minority channel. ($O_M$-2p+Mn-3d) orbital in both spin channels. |
| D | $O_S$ | Ce, Mn | $O_S$-2p orbital mixed with Ce-4f and Mn-3d |
| E | O | Ce | Highly delocalized orbital showing mostly (O-2p+Ce-4f) character. |

*Table 1 The characteristics of the excited-electron giving rise to the salient peaks in the experimental O K-edge spectrum. Mn/O atom without a subscript indicates substantial contribution from both "middle" (M) and "side" (S) atoms.*

The utility of computation in the analysis of x-ray spectroscopy can be appreciated from Table 1 which maps the low-energy spectral peaks to the orbitals of the excited electron. The peaks in region I of the spectrum (see Fig. 8(a)) arise mostly from mixing of O-2$p$ with Mn-3$d$ orbitals while the contribution of Ce-4$f$ is noticed in region II.

A more detailed visual description of the associated orbitals is provided by the relevant isovalue plots of Fig. 8(d)-(m). The orbital plot in panel (d) of Fig. 8 reveals that, in accordance with our theoretical model, the first peak of the $O_M$ K-edge (contributing to peak A) in the up spin channel originates from a final state where the excited electron resides in an orbital with substantial contribution from the 3$d$ orbital of $Mn_M$, which has a net positive spin-polarization. Panel (e) shows that the down spin counterpart essentially involves the 3$d$ orbital of the neighboring $Mn_S$ atom, which has a net negative spin-polarization. Similar to the intense first peak, for $O_M$, the feature at ~530.5 eV (contributing to peak B) can be associated with an orbital with mixed O-2$p$ and Mn-3$d$ character (panel (f)) which, for brevity, has been plotted for the up-spin channel only. However, unlike the orbital associated with the first peak (i.e., panel (d)), this orbital corresponds to the minority spin channel of the relevant Mn atom and consequently has a higher energy, more delocalization and lower intensity. Note that, while both panel(d) and (f) correspond to peaks in the up-spin spectrum, the contributing Mn atom for panel (d) is the $Mn_M$, which has a positive spin-polarization while that for panel (f) is the $Mn_S$, which has a negative spin-polarization. The orbital associated with the higher energy peak at ~532.2 eV (panel (g), peak E), which shows even more delocalization, can be seen to contain appreciable contribution from the Ce-4$f$ orbitals.

Similar to the $O_M$ spectrum, the first peak of the $O_S$ spectrum (contributing to peak A') in the down spin channel originates from mixing of O-2$p$ with the 3$d$ orbital of the neighboring $Mn_S$ atom, which has a negative spin-polarization (panel (h) in Fig. 8). By contrast, for the up spin contributions in the $O_S$ spectrum, as already shown in Fig. 3 and Fig. 7, the lack of exchange interaction pushes the minority (up) spin $Mn_S$-3$d$ orbitals to higher energy, and the first $O_S$ K-edge peak in this spin-channel (at ~529 eV, between A and B), in fact, arises primarily from an unoccupied orbital with heavy contributions from Ce-4$f$, as evident from panel (i) in Fig. 8. This explains the relatively higher energy and lower intensity of this peak, in comparison with the down-spin counterpart. The up-spin peak at ~530.4 eV originates from an orbital similar in characteristics to the one responsible for the peak at ~530.5 eV for the $O_M$ spectrum, as can be seen by noting the similarity between panels (f) and (j).

The higher-energy resultant peak at ~531.8 eV (peak D) contains appreciable contribution from both spin channels of the $O_S$ XAS spectrum. The comparability of the energy and the intensity of these two peaks corresponding to different spin channels can be attributed to the fact that they are associated with unoccupied orbitals containing contributions mostly of $4f$ orbitals of the neighboring Ce atom, which has negligible magnetic moment (panels (k) and (l)).

Panels (d), (e) and (h) reveal that all of the intense low-energy XAS peaks contain substantial contribution from a $3d$ orbital of a Mn atom coordinated with the core-excited oxygen $O^{(x)}$ [note that in panels (d) and (e), this is an $O_M$ atom while in panel (h), this is an $O_S$ atom]. However, the $2p$ contribution of $O^{(x)}$ is noticeably lower than that of other (i.e., not core-excited) O atoms coordinated with the aforementioned Mn. This result is somewhat counterintuitive, since we might expect the core-hole to attract the promoted electron mostly to the $O^{(x)}$ atom. However, as a purely quantum-mechanical effect, the presence of a core-hole would lower the onsite energy of $O^{(x)}$, resulting in a reduction of its relative contribution to the anti-bonding orbitals responsible for these spectral peaks. By contrast, we see the largest $2p$ contribution for the O atom on the opposite side of the Mn atom from the core-excited $O^{(x)}$, due to its farther distance from the core hole [evident in panels (d), (e) and (h)]. Notably, the MBXAS method captures the contributions of both the excited and occupied valence orbitals to these spectral features and leads to strong XAS intensities which would be underestimated with single-particle core-hole approaches.

In short, the O $K$-edge spectrum of the pristine BCM crystal is generated from a non-trivial combination of contribution from the two different spin-channels of the two inequivalent O sites: $O_M$ and $O_S$. As summarized in Table 1, the low-energy features result mostly from promoting the O $1s$ electron to orbitals of mixed O-$2p$ and Mn-$3d$ character, while the presence of Ce-$4f$ is noticed at relatively higher energies.

**(4)     The Reduced BCM Crystal**

As mentioned in the introduction, for the STCH operation, BCM is exposed to high-temperature in an inert atmosphere, resulting in removal of oxygen atoms and it is this reduced phase that takes part in $H_2$ generation by splitting the $H_2O$ molecule. This prompts us to investigate the electronic structure of the thermally-reduced BCM crystal. In particular, owing to the high electronegativity of the O atom, bonding between O and metallic ligands typically involves concentration of electronic population around the former. It is then instructive to probe how the excess electronic population available in the system as a result of the oxygen-removal, arranges itself.  On the experimental front, as shown in Fig. 13, O $K$-edge XAS reveals a marked reduction in the first-peak intensity in the reduced crystal in comparison with the pristine counterpart. A detailed theoretical analysis is required to shed light on the origin of such reduction.

For a vacancy at the $O_M$ site, the formation energy (2.7 eV) is seen to be lower than that for a vacancy at the $O_S$ site (3.3 eV), which indicates a higher probability for oxygen-removal from the $O_M$ site. This conclusion is also supported by experimental XAS studies presented in ref.[20], which reports substantial change in the Mn $L$-near-edge spectrum as a function of temperature, implying a progressive reduction of the Mn atom. On the contrary, no major energy-shift is noted in the Ce $M$-edge spectrum. Therefore, in the rest of the paper, the theoretical treatment of electronic structure and XAS will be limited to a reduced structure with a vacancy at the $O_M$ site.

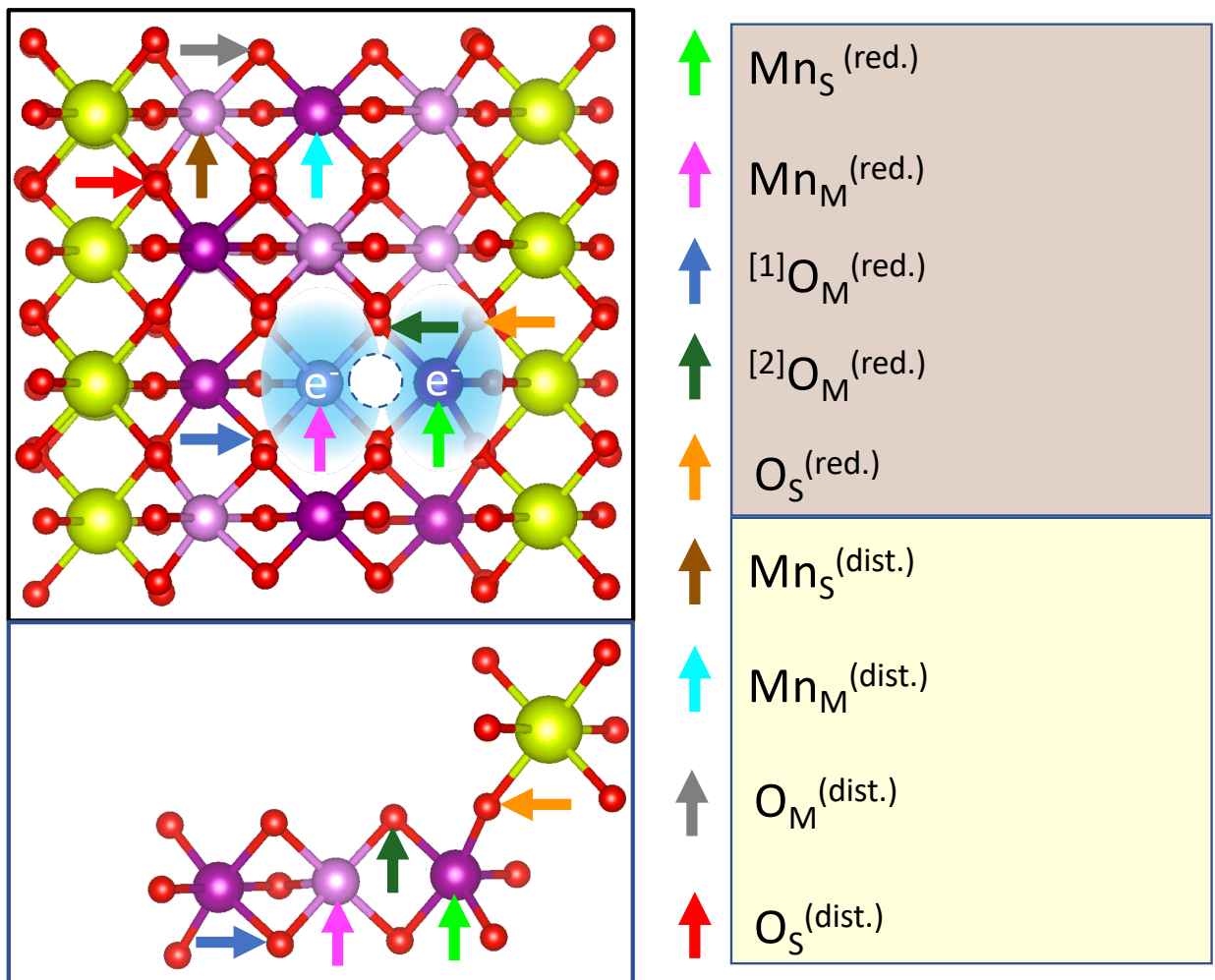

*Fig. 9 Top left panel shows a diagrammatic representation of the reduced crystal structure with the oxygen-vacancy at an $O_M$ site indicated by a dashed circle. The Mn atoms affected most drastically by the removal of $O_M$ are marked by 'e⁻', indicating the presence of excess electronic population (i.e., they are nominally reduced to $Mn^{3+}$). The bottom left panel shows, for clarity, the relative positions of atoms close to the vacancy. Various atoms, which are subjected to further analysis, have been marked by colored arrows with labels in the right-hand-side panel. In the labels, the superscript red., which is short for "reduced" (dist., which is short for "distant") denotes an atom located near (far from) the $O_M$ vacancy.*

The structure of the reduced crystal (provided in the Supplementary Materials) is obtained by relaxing the ionic coordinates in presence of the oxygen vacancy. From Table 2 and Figure 9, it can be seen that the electronic population and the spin-polarization of the atoms nearest the oxygen vacancy (both Mn and O) are noticeably larger than their counterparts in the pristine crystal. We label the

corresponding Mn atoms as Mn$_S$$^{(red.)}$ and Mn$_M$$^{(red.)}$. Due to the vacancy, they are coordinated with 5 (as opposed to 6) oxygen atoms. We would expect that the removal of an O atom would leave behind two electrons on Mn sites. In fact, it is these neighboring Mn atoms, as well as the O atoms coordinated with them which gain electron density (Table 2), and so, we label them as reduced (red.). On the other hand, the more distant (dist.) atoms in the supercell show much smaller deviations in their electron populations. Thus, the excess electronic population available in the system as a result of oxygen-removal, remains local to the site of reduction. A combination of localized orbitals and dielectric screening leads to the effect of the oxygen-vacancy diminishing with increasing distance.

| Atom | Total Population | Spin Up | Spin Down | Polarization |
|---|---|---|---|---|
| **Pristine** | | | | |
| Mn$_S$[C] | 13.4731 | 5.1401 | 8.3330 | −3.1929 |
| Mn$_M$[A] | 13.5225 | 8.3287 | 5.1938 | 3.1350 |
| O$_S$[D] | 6.5212 | 3.2817 | 3.2394 | 0.0423 |
| O$_M$[B] | 6.5680 | 3.2969 | 3.2711 | 0.0258 |
| **Reduced** | | | | |
| Mn$_S$$^{(dist.)}$ | 13.4734 | 5.1408 | 8.3325 | −3.1917 |
| Mn$_M$$^{(dist.)}$ | 13.5217 | 8.3308 | 5.1909 | 3.1398 |
| O$_S$$^{(dist.)}$ | 6.5165 | 3.2822 | 3.2343 | 0.0479 |
| O$_M$$^{(dist.)}$ | 6.5658 | 3.2948 | 3.2710 | 0.0237 |
| Mn$_S$$^{(red.)}$ | 13.5232 | 4.8637 | 8.6595 | −3.7958 |
| Mn$_M$$^{(red.)}$ | 13.5875 | 8.6647 | 4.9228 | 3.7419 |
| O$_S$$^{(red.)}$ | 6.5559 | 3.2970 | 3.2589 | 0.0381 |
| [1]O$_M$$^{(red.)}$ | 6.6015 | 3.3209 | 3.2806 | 0.0403 |
| [2]O$_M$$^{(red.)}$ | 6.6192 | 3.3203 | 3.2988 | 0.0215 |

*Table 2 For various atoms of the pristine and the reduced crystal (column 1), the total electron population (column 2), spin-up population (column 3), spin-down population (column 3) and polarization (column 4) of valence electrons, calculated using Lowdin's scheme of population analysis. For the pristine crystal, the atoms*

*are labeled (A-D) in accordance with Fig. 1 while for the reduced structure, the color-convention used in Fig. 11 is maintained.*

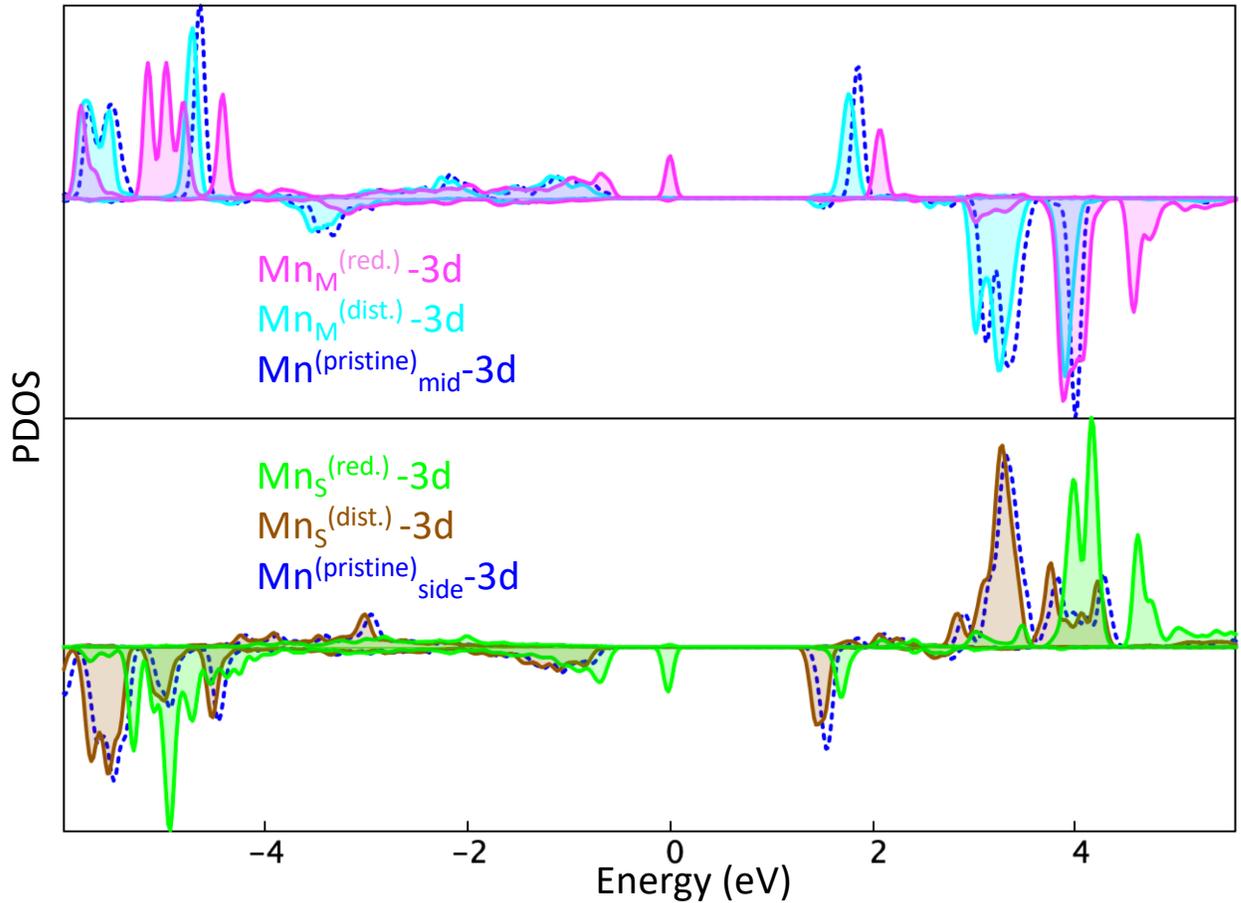

*Fig. 10 Mn-3d PDOS plots contrasting contributions of atoms from the reduced crystal to those of atoms from the pristine one. The top (bottom) panel corresponds to the middle, i.e. M (side, i.e. S) site. The VBM is set to zero.*

Fig. 10 showcases a comparison of the 3$d$ PDOS of different Mn atoms: $Mn_M^{(red.)}$ and $Mn_S^{(red.)}$ ($Mn_M^{(dist.)}$ and $Mn_S^{(dist.)}$) present at the site of reduction (distant from the site of reduction). As expected, the PDOS of the distant atoms, $Mn_M^{(dist.)}$ and $Mn_S^{(dist.)}$, bear noticeable resemblance with their counterparts from the pristine structure, while those of the reduced atoms, $Mn_M^{(red.)}$ and $Mn_S^{(red.)}$, differ appreciably. From Table 2, it can be noted that, owing to their coordination with fewer O atoms, the reduced Mn atoms have higher valence electronic population. This enhanced occupation elevates the unoccupied orbitals to higher energy and shifts their weight off the reduced Mn sites, as evident from the higher energy and lower intensity of the relevant Mn PDOS-peaks, respectively. For the pristine crystal, Fig. 3 reveals that for both of the Mn sites (i.e., $Mn_M$ and $Mn_S$), the unoccupied edge of the PDOS corresponds to the respective majority spin channels. Therefore, after the thermal reduction, the excess electron available to $Mn_M^{(red.)}$ and $Mn_S^{(red.)}$ can be expected to go to the majority spin channel. This is consistent with Table 2, which shows that the absolute values of the polarization of the reduced Mn atoms are higher than those of the pristine counterparts. In Fig. 10, this can be associated with the isolated PDOS contribution in the majority spin channel at the Fermi level (0 eV).

### (5) O *K*-edge XAS of the reduced structure

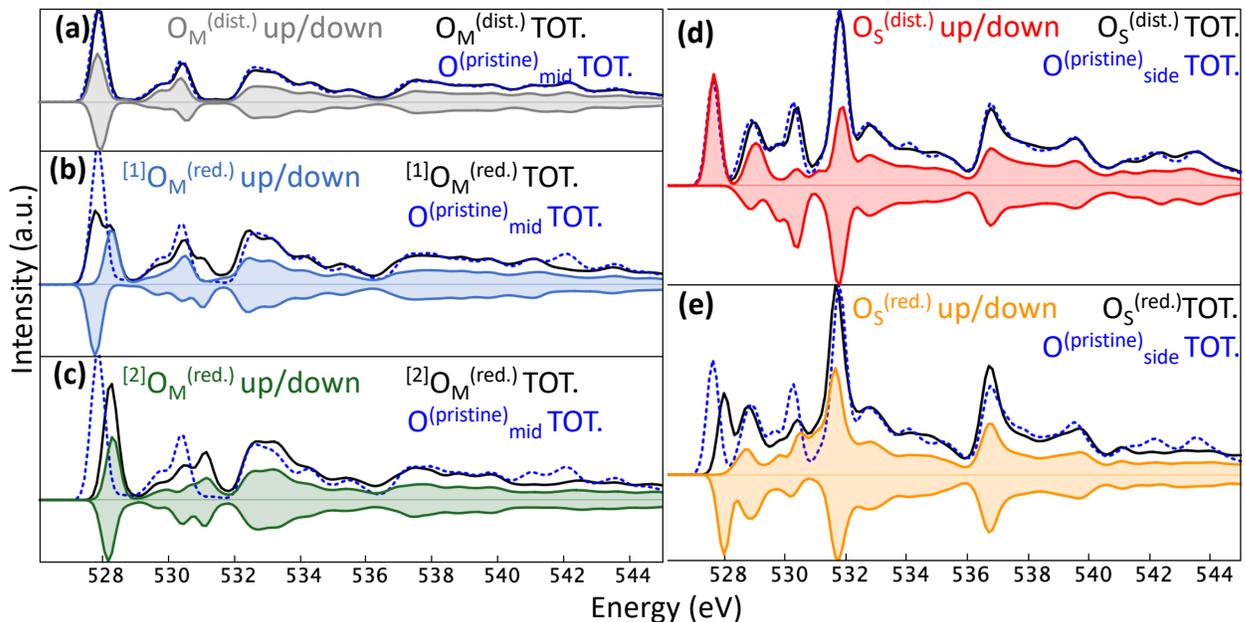

*Fig. 11 O K-edge XAS of different oxygen sites of the reduced crystal. The left (right) panels correspond to $O_M$ ($O_S$) sites. In each panel, the shaded plots show the up and down spin contributions of the relevant O atom, the solid black line shows the spin-summed spectrum, while the broken blue line represents the spectrum of an $O_M$ / $O_S$ atom from the pristine crystal.*

O $K$-edge X-ray absorption spectra corresponding to core-excitation of different inequivalent oxygen sites is displayed in Fig. 11, along with counterparts from the pristine crystal, to facilitate comparison. Once again, the spectra of the distant atoms are very similar to those from the pristine crystal while those of O atoms close to the vacancy-site differ substantially. For $^{[2]}O_M^{(red.)}$, the up and down-spin spectra do not display any drastic asymmetry since both of its neighboring Mn atoms are 5-fold coordinated due to the neighboring vacancy (i.e., they are $Mn_S^{(red.)}$ and $Mn_M^{(red.)}$). However, the increased electron population on these reduced Mn atoms raises the energy of their 3$d$ orbitals, leading to a higher energy-difference between the corresponding Mn-3$d$ and the O-2$p$ levels, effectively making the bonding more ionic (due to decreased orbital mixing). In accordance with the tight-binding formalism, this results in a reduced O-2$p$ contribution to the anti-bonding orbitals (see schematic in Fig. 12). Therefore, the intensity of the resultant $^{[2]}O_M^{(red.)}$ spectrum is lower than the $O_M$ counterpart from the pristine crystal. On the other hand, the $^{[1]}O_M^{(red.)}$ atom is coordinated with two Mn atoms among which, only $Mn_M^{(red.)}$, which has a positive spin-polarization, experiences a major enhancement in electron population. Mixing between $^{[1]}O_M^{(red.)}$-2$p$ and $Mn_M^{(red.)}$-3$d$ orbitals is primarily responsible for the up-spin spectrum of $^{[1]}O_M^{(red.)}$. Therefore, for $^{[1]}O_M^{(red.)}$, the first peak of the up-spin spectrum exhibits a noticeably higher energy and lower intensity compared to the down-spin counterpart. The energy-shift between the first peaks corresponding to these two spin-channels leads to a substantial reduction in the intensity of the resultant peak for $^{[1]}O_M^{(red.)}$, as can be seen by comparing the panels (a) and (b) of Fig. 11. The peak intensity for $O_S^{(red.)}$ is also lowered in comparison with that of pristine $O_S$, although this lowering is less pronounced since, in the pristine as well as in the reduced crystal, only the first peak originates from mixing with Mn orbitals. The second peak, which corresponds to a different spin channel and is much lower in intensity, results mostly from mixing with Ce orbitals and therefore, does not experience any major change upon reduction. The up-spin peak of $^{[1]}O_M^{(red.)}$, the down-spin peak of $O_S^{(red.)}$, and both peaks of $^{[2]}O_M^{(red.)}$ are associated with orbitals resulting from the mixing of O-2$p$ with the 3$d$ orbital of the reduced (i.e., 5 fold coordinated) Mn atom(s). The blueshift of these peaks, with respect to their counterparts in the pristine spectrum is consistent with previous reports[21] and expected from the tight-binding consideration.

As discussed above, for all three O atoms near the site of reduction, i.e., for atoms $^{[1]}O_M^{(red.)}$, $^{[2]}O_M^{(red.)}$ and $O_S^{(red.)}$, the first peak of the spectrum shows a reduction in intensity. Therefore, the resultant O K-edge spectrum of the reduced crystal should exhibit lowering of the first peak intensity, although the exact ratio can be expected to depend on the percentage of oxygen-vacancies formed in the crystal. However, as shown in Fig. 13, at least in qualitative terms, the experimental O *K*-edge spectrum of reduced BCM corroborates the aforementioned conclusion of intensity-lowering.

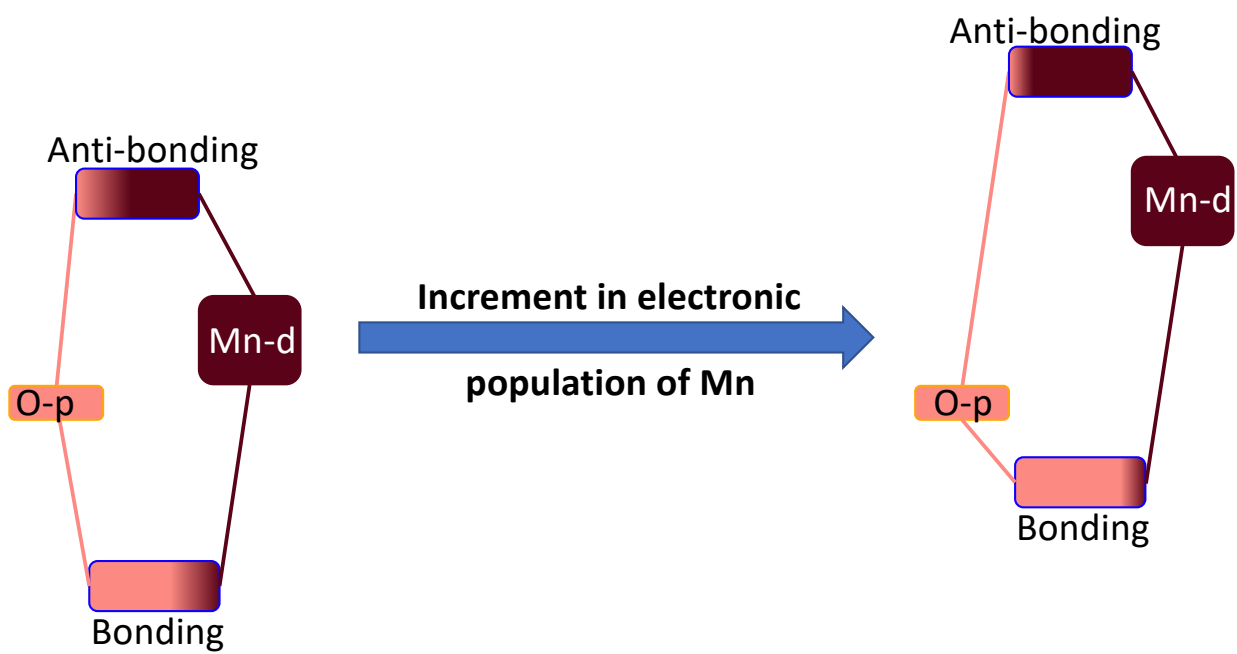

*Fig 12 Schematic diagram showing a reduction in the O-p contribution to the anti-bonding orbital resulting from a higher population of Mn. To facilitate comparison, the O-p level is kept at the same energy. This effect is further exaggerated by the core-excited state, which draws down the O 2p orbital energies and enhances the ionicity, further reducing O character in the antibonding orbital.*

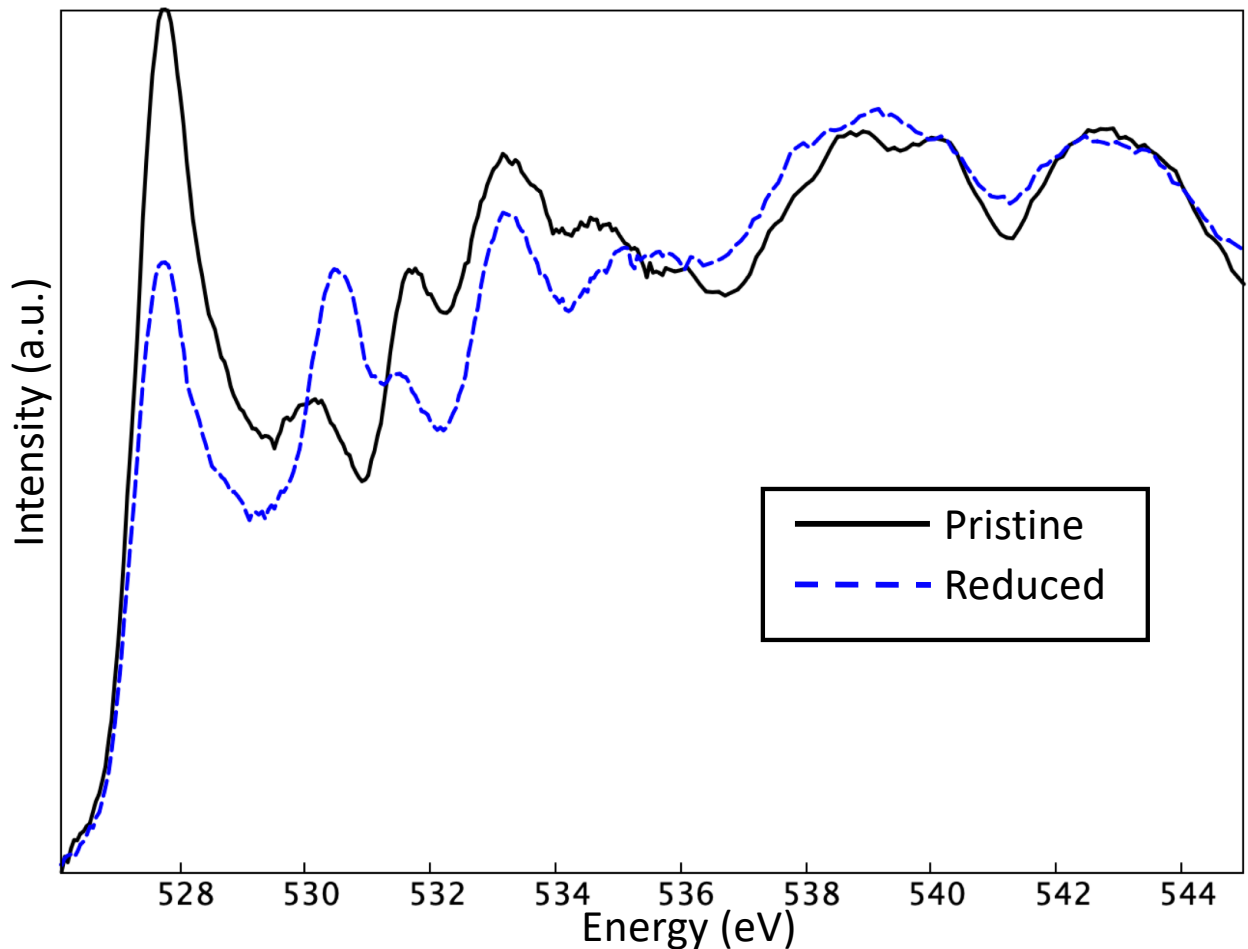

*Fig 13 Experimental O K-edge absorption spectra of pristine and reduced BCM.*

As a consequence of the reduction, in general the electronic population increases in the vicinity of the site of reduction. From the consideration of classical Coulombic interaction, this can be expected to result in a diminished contribution from the local DOS of these atoms to the low-energy conduction orbitals of the system. However, as we have seen in the case of the pristine crystal, quantum mechanical effects associated with orbital-mixing (within the tight-binding framework) can play a crucial role in deciding the character of the final orbital. In order to investigate further how the interplay of the O-2$p$ and Mn-3$d$ orbitals results in a reduction in the peak intensities for the O$^{(red.)}$ atoms, we compare, for the reduced crystal,

A. the PDOS of $Mn_S^{(dist.)}$-3d and $O_S^{(dist.)}$-2p as a consequence of core-exciting the $O_S^{(dist.)}$ atom (note that Fig. 11 indicates that this should be relatively similar to the pristine counterpart).

B. the PDOS of $Mn_S^{(red.)}$-3d and $O_S^{(red.)}$-2p as a consequence of core-excitation of the $O_S^{(red.)}$ atom, against

For the final core-excited state corresponding to case [A], (case[B]), let us denote the KS orbital responsible for generating the peak at 9.94 eV (10.34 eV) as $\phi^A$ ($\phi^B$). Further, let $^{[A]}C[Mn]$ and $^{[A]}C[O]$ ($^{[B]}C[Mn]$ and $^{[B]}C[O]$) denote, respectively, the contribution from the Lowdin-orthogonalized $Mn_S^{(red.)}$-3$d$ and $O_S^{(red.)}$-2$p$ ($Mn_S^{(dist.)}$-3$d$ and $O_S^{(dist.)}$-2$p$) atomic-orbitals to $\phi^A$ ($\phi^B$). Note that the $O_S^{(red.)}$-2$p$ and the $O_S^{(dist.)}$-2$p$ orbitals are responsible for the corresponding first-peaks in the absorption spectra shown in Fig. 11. The computed values of the aforementioned contributions reveal that

$\frac{^{[A]}C[Mn]}{^{[B]}C[Mn]} = 0.86$ while $\frac{^{[A]}C[O]}{^{[B]}C[O]} = 2.3$.

In other words, creation of the oxygen-vacancy results in a lowering of the contribution of the 2$p$ orbital of the core-excited $O_S$ atom at the conduction band-edge. Conversely, the contribution of the $Mn_S$ 3$d$ orbital is enhanced. We note here that the difference in the Lowdin population of the $Mn_S^{(red.)}$ atom in case [B] (i.e., after core-excitation of $O_S^{(dist.)}$) and the $Mn_S^{(dist.)}$ atom in case [A] (i.e., after core-excitation of $O_S^{(red.)}$) is 0.06 while that between the corresponding core-excited oxygens is 0.01. The higher contribution of the 2$p$ orbital of the excited oxygen far from the site of oxygen-vacancy essentially points to the fact that, owing to the higher electronic population of $Mn^{(red.)}$ [see Tab. 2], the difference in the onsite energy between $Mn_S^{(red.)}$-3d and core-excited $O_S^{(red.)}$-2p in case [A] is higher than that between $Mn_S^{(dist.)}$-3d and core-excited $O_S^{(dist.)}$-2p in case [B]. Therefore, in accordance with the tight-binding prescription, in an anti-bonding orbital with mixed $Mn_S$-3$d$ and $O_S$-2$p$ character, the $O_S$-2$p$ contribution would be relatively lower in case [A]. This effect is demonstrated schematically in Fig. 12. The lower $O_S$-2$p$ contribution in case [A] would, in turn, result in a reduction in the intensity of the corresponding XAS peak.

To summarize, a qualitative theoretical analysis combining first-principles calculations and a tight-binding model reveals that the lowering of the first-peak intensity in the reduced crystal originates from a combination of two effects:

1. Lifting of degeneracy between the up and the down spin peaks for $^{[1]}O_M^{(red.)}$, as shown in Fig. 11, and
2. Reduction in the O-2*p* contribution to the unoccupied orbitals, as illustrated schematically in Fig. 12.

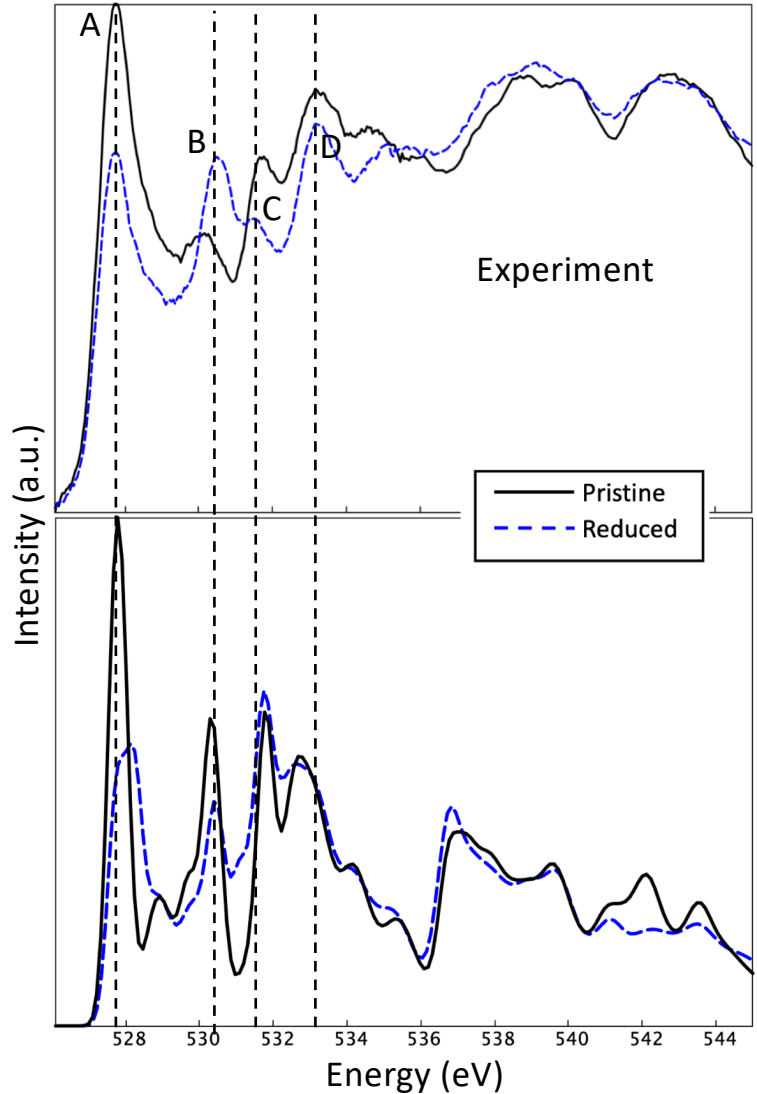

*Fig. 14 Top panel shows the experimental O K-edge spectra of the pristine and the reduced crystal (same as Fig. 13). Bottom panel shows the simulated spectrum of the pristine crystal in black solid line (same as that shown in black line in Fig. 8(b). The broken blue line in the bottom panel shows a model spectrum calculated approximately for the reduced triple. Further details are provided in the text*

Fig. 14 shows, along with the experimental counterparts, the simulated O *K*-edge spectrum for pristine BCM and a model spectrum for the reduced crystal. In the latter, we have accounted, in an approximate manner, only for the O atoms coordinating the reduced triple (see Fig. 9 in the main manuscript). More specifically, this is calculated as $(1/11) \times \left[ 3 \times I(\ O_S^{(dist.)}\ ) + 3 \times I(\ O_S^{(red.)}\ ) + 3 \times I(\ ^{[1]}O_M^{(red.)}\ ) + 2 \times I(\ ^{[2]}O_M^{(red.)}\ ) \right]$, where $I(O_m)$ denotes the spectrum of the $O_m$ atom. Note that the actual spectrum of the reduced crystal will contain contributions from all O atoms present in the crystal, not just the O atoms coordinating the reduced triple. The contributions coming from all O atoms occupying inequivalent sites in the reduced crystal will, in principle, be different. Therefore, the final spectrum will depend, in a complicated manner, not only on the density but also on the distribution of the defects. If the thermal reduction results in removal of two oxygens coordinating the same triple, the spectra would likely be significantly different. Note, for example, that the first-peak of the experimental absorption spectra for the pristine and the reduced crystal are coincident on the energy-axis. On the other hand, in the O $K$-edge EELS spectra reported in Ref.[20], a noticeable blueshift is reported with increasing thermal reduction, especially at 700° C. As discussed in section (5), such a blueshift is expected in the spectral contribution of the $O^{(red.)}$ atoms. Similarly, owing to the change in the anti-bonding orbital as well as spin-splitting, the $O^{(red.)}$ $K$-edge spectra will experience a reduction in the intensity of the first peak. The degree of prominence of such features in the net experimental spectrum will, however, be dictated by the density and distribution of the oxygen vacancies and by the probe-depth and resolution of the experiment. Notably, the measured overall red-shift in the spectral features at and above 536 eV is reproduced in our simulations.

## Conclusion

In conclusion, with a combination of theoretical and experimental studies, in this paper we present a detailed description of the electronic structure of the $BaCe_{0.25}Mn_{0.75}O_{3-\delta}$ crystal in its pristine and reduced form. The electronic structure is probed experimentally with the help of O $K$-edge XAS, which shows a substantial reduction in the first-peak intensity as a result of oxygen-removal. On the theory front, DFT based first-principles calculations are performed to obtain the PDOS and to plot, in real-space, the occupied and the unoccupied orbitals. On the basis of these ab initio results, we propose a simple scheme of orbital-mixing between the O and the ligand orbitals consistent with the tight-binding model and the crystal field theory. Furthermore, using state-of-the-art spectroscopic simulations using the MBXAS method, we provide an in-depth analysis of the O $K$-edge spectrum by associating the spectral peaks with the relevant unoccupied electronic orbitals. We show that the low-energy peaks can mostly be attributed to O-2$p$ and Mn-3$d$ mixing, while the presence of Ce-4$f$ orbitals are observed only at relatively higher energy. For the reduced crystal, our calculations reveal that, owing to dielectric screening, the change in electron-density is heavily localized around the site of reduction. At atomic sites distant from the oxygen-vacancy, the density is practically identical to that of the pristine counterpart. Finally, the simulated spectra replicate the qualitative lowering of the $K$-edge intensity for O atoms close to the vacancy-site. Within the tight-binding formalism, this can be explained in terms of the reduction in the relative contribution of these O atoms to the frontier unoccupied orbitals.

# Supplementary Materials for "Investigating the Electronic Structure of Prospective Water-splitting Oxide BaCe$_{0.25}$Mn$_{0.75}$O$_{3−δ}$ Before and After Thermal Reduction"

Subhayan Roychoudhury§, Sarah Shulda¶, Anuj Goyal¶, Robert Bell¶, Sami Sainio‡, Nicholas Strange‡, James Eujin Park#, Eric N. Coker#, Stephan Lany¶, David Ginley¶, and David Prendergast§

§ The Molecular Foundry, Lawrence Berkeley National Laboratory, Berkeley CA 94720, USA
¶National Renewable Energy Laboratory, Golden, Colorado 80401, USA
‡SLAC National Accelerator Laboratory, Menlo Park, CA 94025
#Sandia National Laboratories, Albuquerque, New Mexico 87185, USA

## Details of XAS Experiment:

X-Ray absorption spectroscopy data was acquired at the Stanford Synchrotron Radiation Lightsource (SSRL). For analysis of the oxygen K-edge, collected on beam line 10-1, a thin layer of BCM powder was spread on carbon tape. Total fluorescence yield spectra were obtained with a silicon diode AXUV100. An oxygen reference spectrum was collected simultaneously with data collection at the oxygen edge for both (reduced and oxidized) samples. The sample data was put on an absolute energy grid for direct comparison by aligning the refence spectra and shifting the sample data correspondingly. Specific to this beamline, an agglomerate of chemical species resides permanently within the beam path for reference collection. Two spectra at three different spots were collected and averaged for each sample. Alignment and averaging were carried out in Athena[1]. Normalization was carried out with Igor Pro (Wavemetrics, Lake Oswego, OR, USA). O1s spectra were base line corrected by subtracting the minimum pre-edge value to bring pre-edge to 0 and then were divided by the average of the region after the main features (after 547eV). In-situ heating experiments at the Mn K-edge and Ce L-edges were carried out on beamline 4-1 using an in-situ chamber fabricated at Sandia National Laboratory. All data was collected in fluorescence using a Ge array fluorescence detector. Three spectra were collected, normalized, aligned, and then averaged for each sample. For normalization, the pre-edge was fit to a line and the extended region fit to a 3-term quadratic polynomial. Reference foils were collected simultaneously with sample data acquisition to enable accurate alignment of all spectra (to place all spectra on an absolute energy grid). Data normalization, alignment, and averaging was carried out in Athena[1].

## Computational Details:

DFT based first-principles calculations of electronic structure are performed with the Quantum Espresso software package. Plane-wave energy cutoff of 35 eV and 280 eV are used for wavefunction and energy, respectively. PBE exchange-correlation functional and Ultrasoft pseudopotentials have been used throughout. In calculations of the x-ray excited states, the core-excited oxygen atom is represented by a modified pseudopotential which encodes the effects of the core-hole. Hubbard U values of 3.9 eV and 5.4 eV are used for the Mn-3$d$ orbitals and the Ce-4$f$ orbitals, respectively. In order to mitigate the effects of the interaction between the periodic images of the core-excited atom, we use a large supercell in our XAS calculations (parameters provided below). Therefore, the self-consistent field calculations are performed at the $k$=0 point only. A uniform (i.e., independent of the incident frequency) broadening of 0.2 eV is used to generate the spectra. Results obtained using other energies for broadening are shown in Fig. S2. The relative alignment of the spectra of different O atoms is determined with the help of the total energies of the FCH states and the energies of the lowest unoccupied orbitals. If, for the FCH state corresponding to core-excitation of the $i$-th ($j$-th) O atom, $E_i$ ($E_j$) is the sum of the total energy and the KS energy of the lowest unoccupied orbital, then the onset-energy for the spectrum of the $i$-th O atom will be higher than that of the $j$-th O atom by ($E_i$ - $E_j$).

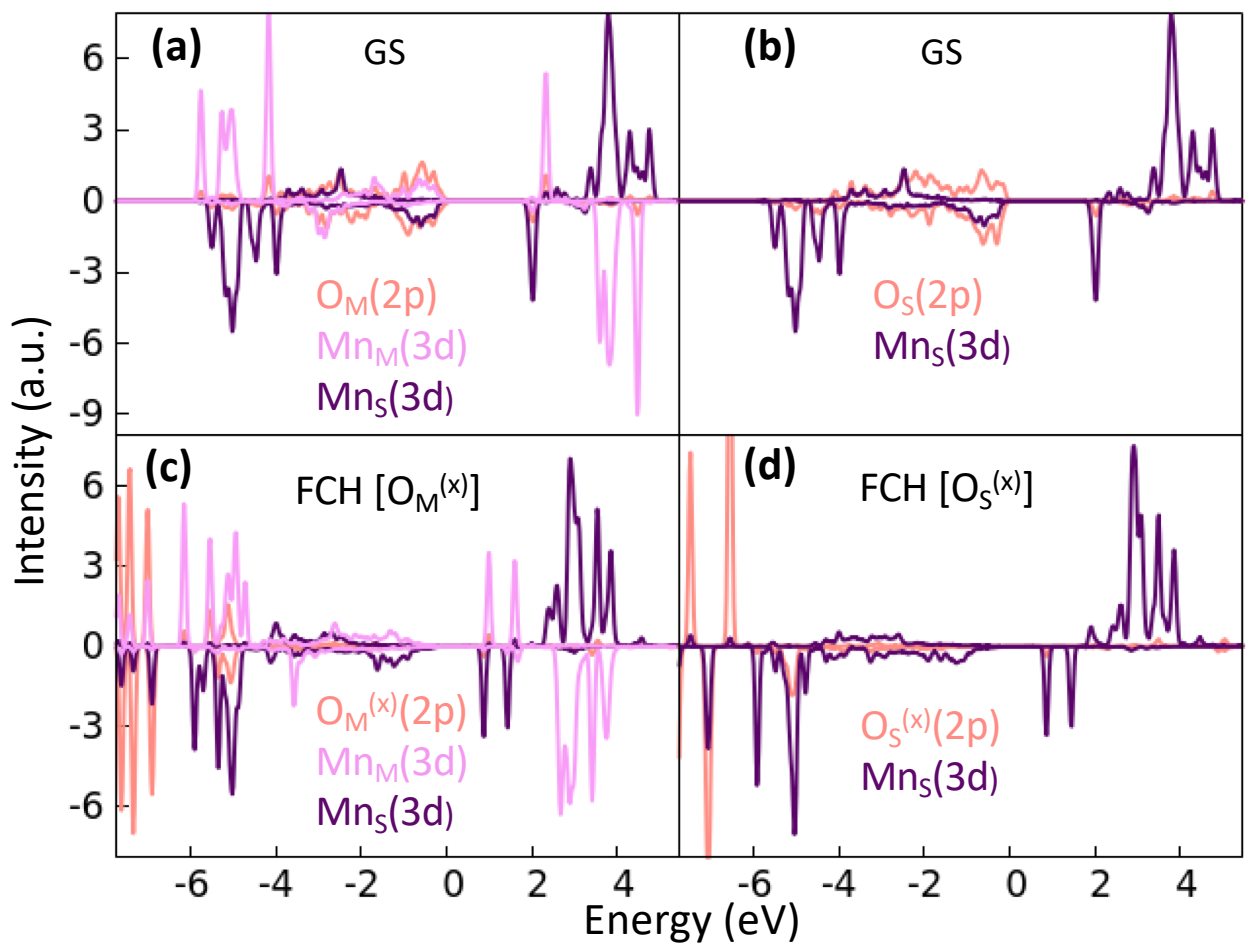

*Fig. S1 Figure showing the PDOS of Mn and O atoms before and after the X-ray absorption process. Panel (a) and panel (b) show, respectively, the GS 2p PDOS of an $O_M$ and an $O_S$ atom along with the 3d PDOS of the neighboring atoms. Panel (c) (panel(d)) shows, for the FCH state obtained by core-exciting an $O_M$ ($O_S$) atom, the 2p PDOS of the core-excited atom $O_M^{(x)}$ ($O_S^{(x)}$), along with the 3d PDOS of the neighboring Mn atoms.*

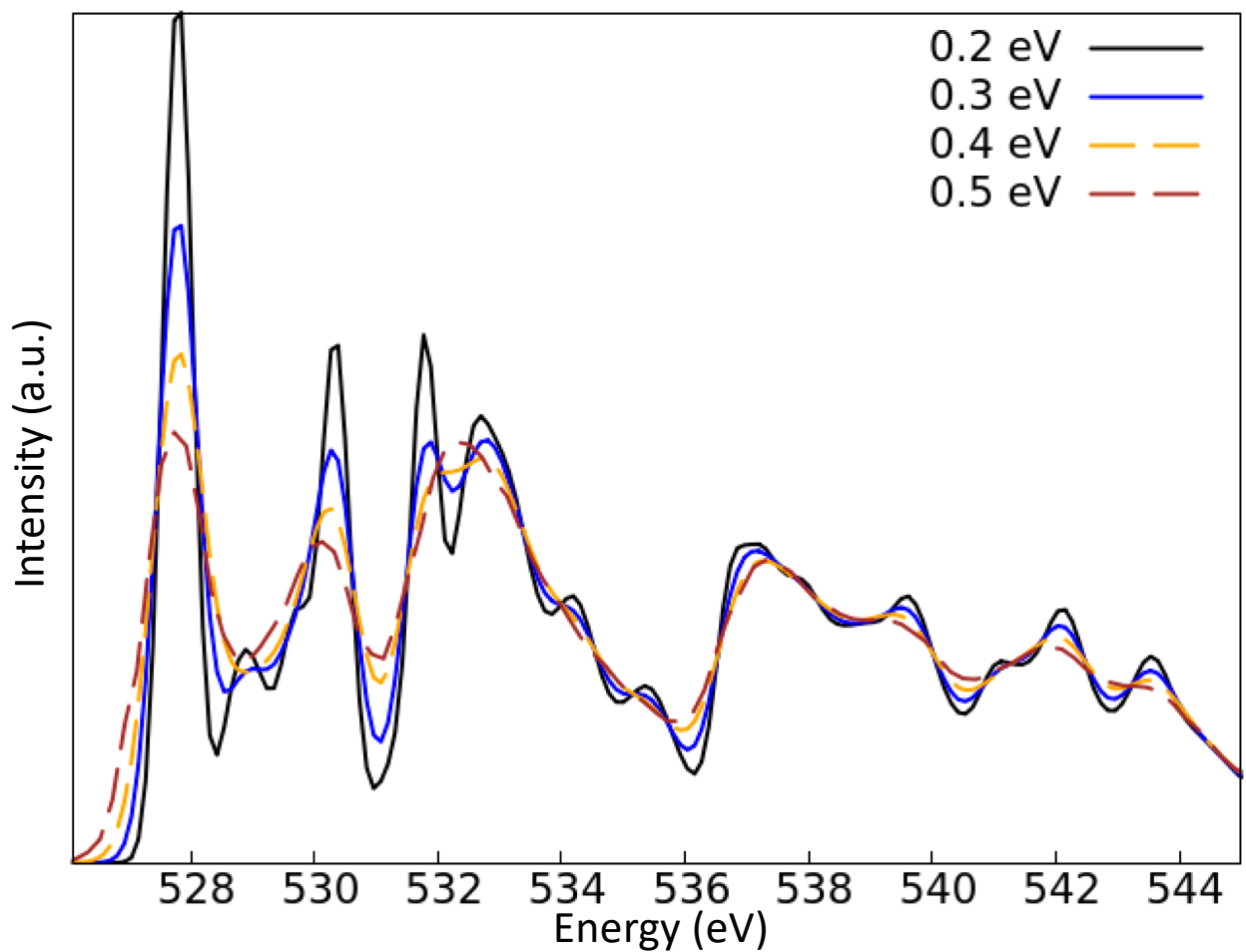

*Fig. S2 O K-edge spectrum of the pristine BCM crystal using different energies for broadening.*

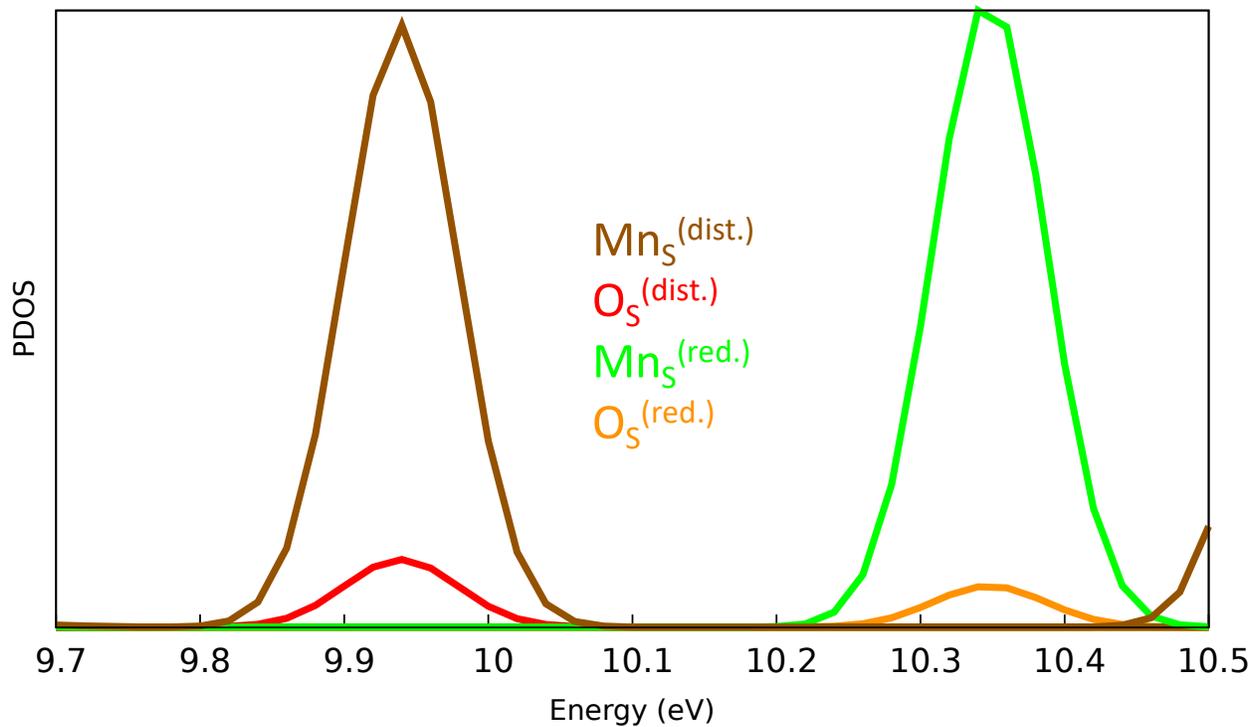

*Fig. S3 The green and orange lines show, at the conduction band edge, the PDOS, respectively, of $Mn_S^{(red.)}$-3d and $O_S^{(red.)}$-2p corresponding to the SCF resulting from core-excitation of the $O_S^{(red.)}$ atom. The brown and the red lines show the conduction band-edge PDOS of $Mn_S^{(dist.)}$-3d and $O_S^{(dist.)}$-2p for the SCF resulting from core-excitation of the $O_S^{(dist.)}$ atom. The $O_S^{(red.)}$-2p orbital generates the first-peak in the $O_S^{(red.)}$ K-edge spectrum shown in Fig. 11 (bottom right panel) while the $O_S^{(dist.)}$-2p orbital is responsible for the $O_S^{(dist.)}$ K-edge first-peak in Fig. 11 (top tight panel).*

**Structure of the Crystal**

Superell Parameters

Lattice Parameter, $a_0$ : 38.5438  a.u.

Crystal axes in units of $a_0$
            a(1) = (  0.500000  -0.866025  -0.000065 )

a(2) = (  0.500041  0.288699  0.000000 )
a(3) = ( -0.166698  0.288729  0.953047 )

Ionic coordinates of the pristine crystal in Angstrom (Mn1 and Mn2 atoms are spin-polarized along opposite directions) :

```
Ba   3.3992982800040   −5.8877571473726    5.9030879974125
Ba   1.6992753014839   −2.9432306153339   15.6224917664166
Ba   8.4988409553240   −2.9435347977983    5.9030879992436
Ba   6.7988179768038    0.0009917342405   15.6224917682477
Ba   8.4984193486792  −14.7196939519691    5.9024215079341
Ba   6.7983963701591  −11.7751674199303   15.6218252769381
Ba  13.5979620239991  −11.7754716023947    5.9024215097652
Ba  11.8979390454790   −8.3309450703559   15.6218252787692
Ba   3.3995090833264    0.0003224297127    5.9034212430673
Ba   1.6994861048063    2.9448489617515   15.6228250120713
Ba   8.4990517586464    2.9445447792871    5.9034212448984
Ba   6.7990287801262    5.8890713113259   15.6228250139024
Ba   8.4986301520016   −8.3316143748837    5.9027547535889
Ba   6.7986071734815   −5.8870878428449   15.6221585225929
Ba  13.5981728273215   −5.8873920253093    5.9027547554200
Ba  11.8981498488014   −2.9428654932706   15.6221585244240
Ba   5.0991316821485   −2.9435103526960    3.8153160382894
Ba   3.3991087036284    0.0010161793427   13.5347198072935
Ba  10.1986743574684    0.0007119968783    3.8153160401205
Ba   8.4986513789483    2.9452385289171   13.5347198091246
Ba  10.1982527508237  −11.7754471572924    3.8146495488110
Ba   8.4982297723035   −8.3309206252537   13.5340533178150
Ba  15.2977954261436   −8.3312248077181    3.8146495506421
Ba  13.5977724476235   −5.8866982756793   13.5340533194664
Ba   5.0993424854709    2.9445692243894    3.8156492839442
Ba   3.3993195069508    5.8890957564281   13.5350530529482
Ba  10.1988851607908    5.8887915739637    3.8156492857753
Ba   8.4988621822707    8.8333181060025   13.5350530547793
Ba  10.1984635541461   −5.8873675802071    3.8149827944658
Ba   8.4984405756259   −2.9428410481683   13.5343865634698
Ba  15.2980062294660   −2.9431452306327    3.8149827962969
Ba  13.5979832509459    0.0013813014061   13.5343865653009
Ba   5.0990564182092   −2.9433798354133    8.2066141932703
Ba   3.3990334396890    0.0011466966254   17.9260179622743
Ba  10.1985990935291    0.0008425141610    8.2066141951014
Ba   8.4985761150090    2.9453690461998   17.9260179641054
Ba  10.1981774868843  −11.7753166400098    8.2059477037918
Ba   8.4981545083642   −8.3307901079710   17.9253514727959
Ba  15.2977201622043   −8.3310942904354    8.2059477056229
Ba  13.5976971836841   −5.8865677583966   17.9253514746270
Ba   5.0992672215316    2.9446997416720    8.2069474389250
Ba   3.3992442430114    5.8892262737108   17.9263512079291
```

```
Ba  10.1988098968515    5.8889220912464    8.2069474407561
Ba   8.4987869183314    8.8334486232852   17.9263512097602
Ba  10.1983882902067   -5.8872370629244    8.2062809494466
Ba   8.4983653116866   -2.9427105308856   17.9256847184506
Ba  15.2979309655267   -2.9430147133500    8.2062809512777
Ba  13.5979079870065    0.0015118186888   17.9256847202818
Ba   3.3995843472658    0.0001919124300    1.5121230880864
Ba   1.6995613687456    2.9447184444688   11.2315268570905
Ba   8.4991270225857    2.9444142620044    1.5121230899175
Ba   6.7991040440656    5.8889407940432   11.2315268589216
Ba   8.4987054159409   -8.8317448921664    1.5114565986080
Ba   6.7986824374208   -5.8872183601276   11.2308603676121
Ba  13.5982480912609   -5.8875225425920    1.5114566004391
Ba  11.8982251127407   -2.9429960105533   11.2308603694432
Ba   3.3993735439434   -5.8878876646553    1.5117898424317
Ba   1.6993505654232   -2.9433611326166   11.2311936114357
Ba   8.4989162192633   -2.9436653150810    1.5117898442628
Ba   6.7988932407432    0.0008612169578   11.2311936132668
Ba   8.4984946126185  -14.7198244692518    1.5111233529532
Ba   6.7984716340984  -11.7752979372130   11.2305271219573
Ba  13.5980372879385  -11.7756021196774    1.5111233547844
Ba  11.8980143094183   -8.8310755876386   11.2305271237884
Ce   0.0000000000000    0.0000000000000    0.0000000000000
Ce  -1.7000229785201    2.9445265320388    9.7194037690040
Ce   5.0995426753199    2.9442223495743    0.0000000018311
Ce   3.3995196967998    5.8887488816131    9.7194037708351
Ce   5.0991210686751   -8.8319368045964   -0.0006664894784
Ce   3.3990980901550   -5.8874102725576    9.7187372795256
Ce  10.1986637439951   -5.8877144550221   -0.0006664876473
Ce   8.4986407654750   -2.9431879229833    9.7187372813567
Ce   5.0993318719975   -2.9438572275110   -0.0003332438237
Ce   3.3993088934774    0.0006693045277    9.7190705251804
Ce  10.1988745473175    0.0003651220633   -0.0003332419926
Ce   8.4988515687973    2.9448916541021    9.7190705270115
Ce  10.1984529406727  -11.7757940321075   -0.0009997333021
Ce   8.4984299621526   -8.8312675000687    9.7184040357019
Ce  15.2979956159926   -8.8315716825331   -0.0009997314710
Ce  13.5979726374725   -5.8870451504943    9.7184040375331
Mn1  1.6997598483999    2.9443744408066    4.8597018854176
Mn1 -0.0002631301202    5.8889009728453   14.5791056544216
Mn1  6.7993025237198    5.8885967903809    4.8597018872487
Mn1  5.0992795451997    8.8331233224197   14.5791056562527
Mn1  6.7988809170751   -5.8875623637899    4.8590353959391
Mn1  5.0988579385549   -2.9430358317511   14.5784391649432
Mn1 11.8984235923950   -2.9433400142155    4.8590353977702
Mn1 10.1984006138749    0.0011865178233   14.5784391667743
Mn1  6.7990941696678    0.0005130610966    7.3894409247774
Mn1  5.0990711911476    2.9450395931354   17.1088446937815
Mn1 11.8986368449877    2.9447354106709    7.3894409266085
Mn1 10.1986138664676    5.8892619427097   17.1088446956126
```

```
Mn1 11.8982152383429    −8.8314237434998     7.3887744352990
Mn1 10.1981922598228    −5.8868972114611    17.1081782043030
Mn1 16.9977579136629    −5.8872013939255     7.3887744371301
Mn1 15.2977349351427    −2.9426748618867    17.1081782061342
Mn1  1.6995465958072    −2.9437009840799     2.3292963565793
Mn1 −0.0004763827130     0.0008255479589    12.0487001255833
Mn1  6.7990892711271     0.0005213654945     2.3292963584104
Mn1  5.0990662926070     2.9450478975332    12.0487001274144
Mn1  6.7986676644823   −11.7756377886763     2.3286298671008
Mn1  5.0986446859622    −8.8311112566375    12.0480336361049
Mn1 11.8982103398023    −8.8314154391020     2.3286298689320
Mn1 10.1981873612821    −5.8868889070632    12.0480336379360
Mn2  1.6995490450775    −2.9437051362788     4.8593686397628
Mn2 −0.0004739334426     0.0008213957600    14.5787724087668
Mn2  6.7990917203974     0.0005172132955     4.8593686415939
Mn2  5.0990687418773     2.9450437453343    14.5787724105979
Mn2  6.7986701137527   −11.7756419408752     4.8587021502844
Mn2  5.0986471352325    −8.8311154088365    14.5781059192884
Mn2 11.8982127890726    −8.8314195913009     4.8587021521155
Mn2 10.1981898105525    −5.8868930592621    14.5781059211195
Mn2  1.6997622976702     2.9443702886076     7.3897741686011
Mn2 −0.0002606808499     5.8889968206464    17.1091779376051
Mn2  6.7993049729902     5.8885926381820     7.3897741704322
Mn2  5.0992819944700     8.8331191702207    17.1091779394362
Mn2  6.7988833663454    −5.8875665159888     7.3891076791227
Mn2  5.0988603878252    −2.9430399839500    17.1085114481267
Mn2 11.8984260416653    −2.9433441664144     7.3891076809538
Mn2 10.1984030631452     0.0011823656243    17.1085114499578
Mn2  6.7988784678047    −5.8875582115909     2.3289631127556
Mn2  5.0988554892846    −2.9430316795521    12.0483668817597
Mn2 11.8984211431247    −2.9433358620166     2.3289631145867
Mn2 10.1983981646045     0.0011906700222    12.0483668835908
Mn2 11.8979995364799   −14.7194950161873     2.3282966232772
Mn2 10.1979765579597   −11.7749684841486    12.0477003922812
Mn2 16.9975422117998   −11.7752726666130     2.3282966251083
Mn2 15.2975192332797    −8.8307461345742    12.0477003941123
0   6.0518685395084    −4.5936989226439     6.0810425107902
0   4.3518455609883    −1.6491723906051    15.8004462797943
0  11.1514112148284    −1.6494765730695     6.0810425126213
0   9.4513882363082     1.2950499589693    15.8004462816254
0  11.1509896081836   −13.4256357272403     6.0803760213118
0   9.4509666296634   −10.4811091952015    15.7997797903158
0  16.2505322835035   −10.4814133776659     6.0803760231429
0  14.5505093049834    −7.5368868456272    15.7997797921469
0   6.0520695928534    −1.2931696830465     6.0816616470464
0   4.3520466143333     1.6513568489923    15.8010654160505
0  11.1516122681734     1.6510526665279     6.0816616488775
0   9.4515892896532     4.5955791985667    15.8010654178816
0  11.1511906615286   −10.1251064876429     6.0809951575680
0   9.4511676830084    −7.1805799556041    15.8003989265720
```

```
O  16.2507333368485   −7.1808841380685    6.0809951593991
O  14.5507103583284   −4.2363576060298   15.8003989284031
O   3.1936366615992    2.9444711569437    6.0819948916698
O   1.4936136830791    5.8889976889825   15.8013986606739
O   8.2931793369192    5.8886935065181    6.0819948935009
O   6.5931563583990    8.8332200385568   15.8013986625050
O   8.2927577302744   −5.8874656476527    6.0813284021914
O   6.5927347517543   −2.9429391156139   15.8007321711954
O  13.3923004055943   −2.9432432980784    6.0813284040225
O  11.6922774270742    0.0012832339604   15.8007321730265
O   0.9527377208559    1.6506875444646    6.0819948908701
O  −0.7472852576643    4.5952140765033   15.8013986598741
O   6.0522803961758    4.5949098940389    6.0819948927012
O   4.3522574176557    7.5394364260777   15.8013986617052
O   6.0518587895310   −7.1812492601318    6.0813284013917
O   4.3518358110109   −4.2367227280931   15.8007321703957
O  11.1514014648510   −4.2370269105575    6.0813284032228
O   9.4513784863308   −1.2925003785187   15.8007321722268
O   3.1934258582768   −2.9436084201417    6.0816616460151
O   1.4934028797567    0.0009181118971   15.8010654150191
O   8.2929685335968    0.0006139294327    6.0816616478462
O   6.5929455505766    2.9451404614714   15.8010654168502
O   8.2925469269520  −11.7755452247381    6.0809951565366
O   6.5925239484319   −8.8310186926993   15.8003989255407
O  13.3920896022719   −8.8313228751637    6.0809951583677
O  11.6920666237518   −5.8867963431250   15.8003989273718
O   0.9525366675109   −1.6498416951328    6.0813757546139
O  −0.7474863110093    1.2946848369059   15.8007795236179
O   6.0520793428308    1.2943806544415    6.0813757564450
O   4.3520563643107    4.2389071864803   15.8007795254490
O   6.0516577361860  −10.4817784997292    6.0807092651355
O   4.3516574576659   −7.5372519676905   15.8001130341395
O  11.1512004115060   −7.5375561501549    6.0807092669666
O   9.4511774329858   −4.5930296181161   15.8001130359706
O   2.4467722259665    1.6505109996606    3.6376947705665
O   0.7467492474464    4.5950375316993   13.3570985395705
O   7.5463149012865    4.5947333492349    3.6376947723976
O   5.8462919227663    7.5392598812737   13.3570985414016
O   7.5458932946417   −7.1814258049358    3.6370282810881
O   5.8458703161215   −4.2368992728971   13.3564320500921
O  12.6454359699616   −4.2372034553615    3.6370282829192
O  10.9454129914415   −1.2926769233227   13.3564320519232
O   5.3052149071981    0.0004204971584    3.6370756353416
O   3.6051919286780    2.9449470291971   13.3564794043457
O  10.4047575825180    2.9446428467327    3.6370756371727
O   8.7047346039979    5.8891693787715   13.3564794061768
O  10.4043359758733   −8.8315163074380    3.6364091458632
O   8.7043129973531   −5.8869897753993   13.3558129148673
O  15.5038786511932   −5.8872939578637    3.6364091476943
O  13.8038556726731   −2.9427674258249   13.3558129166984
```

| | | | |
|---|---|---|---|
| 0 | 7.5459030446191 | −4.5938754674479 | 3.6367423904866 |
| 0 | 5.8458800660989 | −1.6493489354091 | 13.3561461594907 |
| 0 | 12.6454457199390 | −1.6496531178735 | 3.6367423923177 |
| 0 | 10.9454227414189 | 1.2948734141652 | 13.3561461613218 |
| 0 | 12.6450241132942 | −13.4258122720443 | 3.6360759010082 |
| 0 | 10.9450011347741 | −10.4812857400055 | 13.3554796700122 |
| 0 | 17.7445667886142 | −10.4815899224699 | 3.6360759028393 |
| 0 | 16.0045438100940 | −7.5370633904312 | 13.3554796718433 |
| 0 | 5.3050041038757 | −5.8876590799270 | 3.6367423896869 |
| 0 | 3.6049811253556 | −2.9431325478882 | 13.3561461586909 |
| 0 | 10.4045467791956 | −2.9434367303527 | 3.6367423915180 |
| 0 | 8.7045238006755 | 0.0010898016861 | 13.3561461605220 |
| 0 | 10.4041251725509 | −14.7195958845234 | 3.6360759002084 |
| 0 | 8.7041021940307 | −11.7750693524847 | 13.3554796692125 |
| 0 | 15.5036678478708 | −11.7753735349491 | 3.6360759020396 |
| 0 | 13.8036448693507 | −8.8308470029103 | 13.3554796710436 |
| 0 | 2.4465711726215 | −1.6500182399368 | 3.6370756343103 |
| 0 | 0.7465481941014 | 1.2945082921019 | 13.3564794033143 |
| 0 | 7.5461138479415 | 1.2942041096375 | 3.6370756361414 |
| 0 | 5.8460908694213 | 4.2387306416763 | 13.3564794051454 |
| 0 | 7.5456922412967 | −10.4819550445333 | 3.6364091448319 |
| 0 | 5.8456692627765 | −7.5374285124945 | 13.3558129138359 |
| 0 | 12.6452349166166 | −7.5377326949589 | 3.6364091466630 |
| 0 | 10.9452119380965 | −4.5932061629201 | 13.3558129156670 |
| 0 | 7.5461040979641 | −1.2933462278505 | 3.6373615267428 |
| 0 | 5.8460811194439 | 1.6511803041883 | 13.3567652957469 |
| 0 | 12.6456467732840 | 1.6508761217239 | 3.6373615285739 |
| 0 | 10.9456237947639 | 4.5954026537627 | 13.3567652975780 |
| 0 | 12.6452251666392 | −10.1252830324469 | 3.6366950372644 |
| 0 | 10.9452021881191 | −7.1807565004081 | 13.3560988062684 |
| 0 | 17.7447678419592 | −7.1810606828725 | 3.6366950390955 |
| 0 | 16.0047448634390 | −4.2365341508338 | 13.3560988080995 |
| 0 | 2.4964507145836 | 4.3239924294312 | 8.4109867367717 |
| 0 | 0.7964277360635 | 7.2685189614700 | 18.1303905057757 |
| 0 | 7.5959933899036 | 7.2682147790055 | 8.4109867386028 |
| 0 | 5.8959704113834 | 10.2127413110443 | 18.1303905076068 |
| 0 | 7.5955717832588 | −4.5079443751652 | 8.4103202472933 |
| 0 | 5.8955488047386 | −1.5634178431265 | 18.1297240162973 |
| 0 | 12.6951144585787 | −1.5637220255909 | 8.4103202491244 |
| 0 | 10.9950914800586 | 1.3808045064479 | 18.1297240181284 |
| 0 | 7.5952984046644 | −1.3785530969637 | 8.4106921260847 |
| 0 | 5.8952754261442 | 1.5659734350750 | 18.1300958950887 |
| 0 | 12.6948410799843 | 1.5656692526106 | 8.4106921279158 |
| 0 | 10.9948181014642 | 4.5101957846494 | 18.1300958969199 |
| 0 | 12.6944194733395 | −10.2104899015602 | 8.4100256366063 |
| 0 | 10.9943964948194 | −7.2659633695214 | 18.1294294056103 |
| 0 | 17.7939621486595 | −7.2662675519858 | 8.4100256384374 |
| 0 | 16.0039391701393 | −4.3217410199470 | 18.1294294074414 |
| 0 | 0.1066303044843 | 2.9442289661446 | 8.4109867359188 |
| 0 | −1.5933926740358 | 5.8887554981834 | 18.1303905049229 |
| 0 | 5.2061729798043 | 5.8884513157190 | 8.4109867377499 |
| 0 | 3.5061500012841 | 8.3329778477577 | 18.1303905067540 |
| 0 | 5.2057513731595 | −5.8877078384518 | 8.4103202464404 |
| 0 | 3.5057283946394 | −2.9431813064130 | 18.1297240154444 |
| 0 | 10.3052940484794 | −2.9434854888774 | 8.4103202482715 |
| 0 | 8.6052710699593 | 0.0010410431613 | 18.1297240172755 |
| 0 | 5.2059621764819 | 0.0003717386336 | 8.4106534920952 |
| 0 | 3.5059391979618 | 2.9448982706724 | 18.1300572610992 |
| 0 | 10.3055048518018 | 2.9445940882079 | 8.4106534939263 |
| 0 | 8.6054818732817 | 5.8891206202467 | 18.1300572629303 |
| 0 | 10.3050832451570 | −8.8315650659628 | 8.4099870026167 |
| 0 | 8.6050602666369 | −5.8870385339241 | 18.1293907716208 |
| 0 | 15.4046259204770 | −5.8873427163885 | 8.4099870044478 |
| 0 | 13.7046029419568 | −2.9428161843497 | 18.1293907734519 |
| 0 | 2.4959665326668 | 1.5653041305473 | 8.4110253699084 |
| 0 | 0.7959435541467 | 4.5098306625861 | 18.1304291389124 |
| 0 | 7.5955092079868 | 4.5095264801216 | 8.4110253717395 |
| 0 | 5.8954862294666 | 7.4540530121604 | 18.1304291407435 |
| 0 | 7.5950876013420 | −7.2666326740491 | 8.4103588804299 |
| 0 | 5.8950646228218 | −4.3221061420103 | 18.1297626494340 |
| 0 | 12.6946302766619 | −4.3224103244748 | 8.4103588822610 |
| 0 | 10.9946072981418 | −1.3778837924360 | 18.1297626512651 |
| 0 | 2.4962399112612 | −1.5640871476541 | 8.4106534911169 |
| 0 | 0.7962169327411 | 1.3804393843846 | 18.1300572601210 |
| 0 | 7.5957825865811 | 1.3801352019202 | 8.4106534929480 |
| 0 | 5.8957596080610 | 4.3246617339590 | 18.1300572619521 |
| 0 | 7.5953609799363 | −10.3960239522505 | 8.4099870016385 |
| 0 | 5.8953380014162 | −7.4514974202118 | 18.1293907706425 |
| 0 | 12.6949036552563 | −7.4518010026762 | 8.4099870034696 |
| 0 | 10.9948806767361 | −4.5072750706374 | 18.1293907724736 |
| 0 | 8.3920104609906 | −5.8874168891279 | 1.3077505454379 |
| 0 | 6.6919874824705 | −2.9428903570891 | 11.0271543144419 |
| 0 | 13.4915531363105 | −2.9431945395536 | 1.3077505472690 |
| 0 | 11.7915301577904 | 0.0013319924852 | 11.0271543162730 |
| 0 | 13.4911315296658 | −14.7193536937243 | 1.3070840559595 |
| 0 | 11.7911085511456 | −11.7748271616856 | 11.0264878249635 |
| 0 | 18.5906742049857 | −11.7751313441500 | 1.3070840577906 |
| 0 | 16.8906512264656 | −8.8306048121112 | 11.0264878267946 |
| 0 | 0.9033423608106 | −1.5646348260196 | 1.3080451552720 |
| 0 | −0.7966806177096 | 1.3798917060192 | 11.0274489242760 |
| 0 | 6.0028850361305 | 1.3795875235548 | 1.3080451571031 |
| 0 | 4.3028620576104 | 4.3241140555936 | 11.0274489261071 |
| 0 | 6.0024634294857 | −10.3965716306160 | 1.3073786657936 |
| 0 | 4.3024404509656 | −7.4520450985772 | 11.0267824347976 |
| 0 | 11.1020061048057 | −7.4523492810416 | 1.3073786676247 |
| 0 | 9.4019831262855 | −4.5078227490029 | 11.0267824366287 |
| 0 | 6.0021900508914 | −7.2671803524146 | 1.3077505445850 |
| 0 | 4.3021670723712 | −4.3226538203758 | 11.0271543135891 |
| 0 | 11.1017327262113 | −4.3229580028402 | 1.3077505464161 |
| 0 | 9.4017097476912 | −1.3784314708014 | 11.0271543154202 |

```
O   11.1013111195665   −16.0991171570110    1.3070840551066
O    9.4012881410464   −13.1545906249722   11.0264878241106
O   16.2008537948865   −13.1548948074366    1.3070840569377
O   14.5008308163663   −10.2103682753978   11.0264878259417
O    6.0026742328081    −4.5084920535306    1.3077119114483
O    4.3026512542880    −1.5639655214918   11.0271156804524
O   11.1022169081281    −1.5642697039563    1.3077119132794
O    9.4021939296079     1.3802568280825   11.0271156822835
O   11.1017953014833   −13.3404288581270    1.3070454219699
O    9.4017723229631   −10.3959023260882   11.0264491909740
O   16.2013379768032   −10.3962065085527    1.3070454238010
O   14.5013149982831    −7.4516799765139   11.0264491928051
O    6.0024008542137    −1.3791007753291    1.3080837902398
O    4.3023778756936     1.5654257567097   11.0274875592438
O   11.1019435295337     1.5651215742452    1.3080837920709
O    9.4019205510135     4.5096481062840   11.0274875610749
O   11.1015219228889   −10.2110375799255    1.3074173007614
O    9.4014989443687    −7.2665110478867   11.0268210697654
O   16.2010645982088    −7.2668152303512    1.3074173025925
O   14.5010416196887    −4.3222886983124   11.0268210715965
O    3.2926785889931    −2.9435596616169    1.3080837892615
O    1.5926556104729     0.0009668704219   11.0274875582656
O    8.3922212643130     0.0006626879575    1.3080837910927
O    6.6921982857929     2.9451892199962   11.0274875600967
O    8.3917996576682   −11.7754964662133    1.3074172997831
O    6.6917766791481    −8.8309699341745   11.0268210687872
O   13.4913423329882    −8.8312741166390    1.3074173016142
O   11.7913193544680    −5.8867475846002   11.0268210706183
```

```
Ba    3.4040895190061    −5.8645513210539    5.9088414399699
Ba    1.7040665404859    −2.9200247890151   15.6282452089740
Ba    8.5032105876812   −14.6964881256503    5.9081749504915
Ba    6.8031876091611   −11.7519615936115   15.6275787194956
Ba    3.3178615110375     0.0685691035000    5.8822084861659
Ba    1.6178385325173     3.0130956355388   15.6016122551699
Ba    8.4169825797126    −8.7633677010964    5.8815419966875
Ba    6.7169596011925    −5.8188411690576   15.6009457656915
Ba    5.0784088649930    −2.9663778313410    3.6793491705317
Ba    3.3783858864729    −0.0218512993022   13.3987529395358
Ba   10.1775299336682   −11.7983146359374    3.6786826810533
Ba    8.4775069551480    −8.8537881038986   13.3980864500573
Ba    5.1319315780046     2.9256405405172    3.7896500451257
Ba    3.4319085994845     5.8701670725559   13.5090538141298
Ba   10.2310526466798    −5.9062962640793    3.7889835556473
```

```
Ba   8.5310296681596    -2.9617697320405    13.5083873246514
Ba   5.0556400633749    -3.0145395684981     8.3745257190662
Ba   3.3556170848548    -0.0700130364594    18.0939294880703
Ba  10.1547611320501   -11.8464763730946     8.3738592295878
Ba   8.4547381535299    -8.9019498410558    18.0932629985918
Ba   5.1206036048368     2.9262833236750     8.2262294990924
Ba   3.4205806263167     5.8708098557138    17.9456332680965
Ba  10.2197246735120    -5.9056534809214     8.2255630096140
Ba   8.5197016949918    -2.9611269488827    17.9449667786181
Ba   3.4231482612154     0.0129761209801     1.5232703971657
Ba   1.7231252826953     2.9575026530189    11.2426741661697
Ba   8.5222693298906    -8.8189606836163     1.5226039076873
Ba   6.8222463513704    -5.8744341515775    11.2420076766913
Ba  13.5928633772548     0.0074040315938     1.5019560279765
Ba  11.8928403987347     2.9519305636326    11.2213597969805
Ba  18.6919844459300    -8.8245327730026     1.5012895384981
Ba  16.9919614674098    -5.8800062409638    11.2206933075021
Ba   8.5950761146876    -2.9919241216079     5.8941707699720
Ba   6.8950531361674    -0.0473975895691    15.6135745389761
Ba  13.6941971833627   -11.8238609262043     5.8935042804936
Ba  11.9941742048426    -8.8793343941655    15.6129080494977
Ba   8.5183723611846     2.9614089366863     5.8964536811389
Ba   6.8183493826645     5.9059354687250    15.6158574501429
Ba  13.6174934298598    -5.8705278679102     5.8957871916605
Ba  11.9174704513396    -2.9260013358714    15.6151909606645
Ba  10.2004668910093     0.0262023754860     3.8158049179340
Ba   8.5004439124892     2.9707289075248    13.5352086869380
Ba  15.2995879596845    -8.8057344291104     3.8151384284556
Ba  13.5995649811643    -5.8612078970716    13.5345421974596
Ba  -0.0090112759983     0.0161983434607     3.8021901139515
Ba  -1.7090342545185     2.9607248754995    13.5215938829555
Ba   5.0901097926768    -8.8157384611357     3.8015236244721
Ba   3.3900868141567    -5.8712119290969    13.5209273934771
Ba  10.1829124240489     0.0043994050102     8.1959829834398
Ba   8.4828894455287     2.9489259370490    17.9153867524439
Ba  15.2820334927240    -8.8275373995862     8.1953164939614
Ba  13.5820105142039    -5.8830108675474    17.9147202629654
Ba  10.1835002637261     5.8806382723380     8.2041582660053
Ba   8.4834772852059     8.8251648043767    17.9235620350094
Ba  15.2826213324012    -2.9512985322585     8.2034917765269
Ba  13.5825983538811    -0.0067720002197    17.9228955455309
Ba   8.4991362103609     2.9463798242817     1.5170712772442
Ba   6.7991132318408     5.8909063563205    11.2364750462483
Ba  13.5982572790361    -5.8855569803147     1.5164047877658
Ba  11.8982343005159    -2.9410304482759    11.2358085567698
Ba   8.5204778274776    -2.8938461208796     1.5623727125110
Ba   6.8204548489575     0.0506804111592    11.2817764815150
Ba  13.6195988961528   -11.7257829254760     1.5617062230325
Ba  11.9195759176327    -8.7812563934372    11.2811099920366
Ce  10.1904409535025     5.8675456206648     0.0053540376008
```

```
Ce   8.4904179749824    8.8120721527035     9.7247578066049
Ce  15.2895620221777   -2.9643911839316     0.0046875481224
Ce  13.5895390436576   -0.0198646518929     9.7240913171264
Ce   3.3974960511682   -0.0016035229482     9.6887710684674
Ce   1.6974730726481    2.9429230090906    19.4081748374714
Ce   8.4966171198434   -8.8335403275446     9.6881045789890
Ce   6.7965941413232   -5.8890137955058    19.4075083479930
Ce   8.4925329900939   -2.9474050072702     9.6989500904609
Ce   6.7925100115738   -0.0028784752314    19.4183538594649
Ce  13.5916540587691  -11.7793418118666     9.6982836009824
Ce  11.8916310802489   -8.8348152798279    19.4176873699865
Ce  10.2083396342900    0.0179591231985     0.0241394416810
Ce   8.5083166557698    2.9624856552373     9.7435432106850
Ce  15.3074607029651   -8.8139776813979     0.0234729522026
Ce  13.6074377244450   -5.8694511493591     9.7428767212066
Mn1  6.7942213973146   -5.8474760341253     4.8549203875392
Mn1  5.0941984187944   -2.9029495020865    14.5743241565433
Mn1 11.8933424659897  -14.6794128387217     4.8542538980608
Mn1 10.1933194874696  -11.7348863066829    14.5736576670649
Mn1  6.7912894739824   -0.0172173067702     7.4192110384332
Mn1  5.0912664954623    2.9273092252686    17.1386148074373
Mn1 11.8904105426576   -8.8491541113666     7.4185445489548
Mn1 10.1903875641374   -5.9046275793278    17.1379483179589
Mn1  1.7206859283224   -2.9138975756940     2.3480077961458
Mn1  0.0206629498023    0.0306289563448    12.0674115651498
Mn1  6.8198069969975  -11.7458343802904     2.3473413066673
Mn1  5.1197840184774   -8.8013078482516    12.0667450756714
Mn1  6.8164613655725    5.8877280669186     4.8572869252217
Mn1  5.1164383870524    8.8322545989574    14.5766906942258
Mn1 11.9155824342477   -2.9442087376778     4.8566204357433
Mn1 10.2155594557275    0.0003177943610    14.5760242047473
Mn1 11.8885284531816    2.9332923492647     7.4065109086546
Mn1 10.1885054746614    5.8778188813035    17.1259146776586
Mn1 16.9876495218567   -5.8986444553317     7.4058444191761
Mn1 15.2876265433366   -2.9541179232929    17.1252481881802
Mn1  6.8130600074064    0.0292125819041     2.3370442435857
Mn1  5.1130370288862    2.9737391139429    12.0564480125897
Mn1 11.9121810760815   -8.8027242226923     2.3363777541072
Mn1 10.2121580975614   -5.8581976906536    12.0557815231113
Mn2  1.7287594246200   -2.9173257043775     4.8784321553607
Mn2  0.0287364460999    0.0272008276613    14.5978359243647
Mn2  6.8278804932952  -11.7492625089739     4.8777656658822
Mn2  5.1278575147750   -8.8047359769352    14.5971694348863
Mn2  1.6912628440275    2.9414190300235     7.3885104220945
Mn2 -0.0087601344926    5.8859455620623    17.1079141910985
Mn2  6.7903839127027   -5.8905177745729     7.3878439326161
Mn2  5.0903609341825   -2.9459912425341    17.1072477016201
Mn2  6.8070200511077   -5.8658842224715     2.3246855720501
Mn2  5.1069970725875   -2.9213576904327    12.0440893410542
Mn2 11.9061411197828  -14.6978210270679     2.3240190825717
```

```
Mn2  10.2061181412627   -11.7532944950291    12.0434228515757
Mn2   6.8280681010776     0.0480381338381     4.8472432749950
Mn2   5.1280451225574     2.9925646658769    14.5666470439990
Mn2  11.9271891697527    -8.7838986707583     4.8465767855166
Mn2  10.2271661912326    -5.8393721387195    14.5659805545206
Mn2   6.8007733673805     5.8748307777536     7.3867700727171
Mn2   5.1007503888604     8.8193573097923    17.1061738417211
Mn2  11.8998944360557    -2.9571060268428     7.3861035832386
Mn2  10.1998714575355    -0.0125794948041    17.1055073522427
Mn2  11.9071226993696    -2.9259170924913     2.3293666533667
Mn2  10.2070997208495     0.0186094395475    12.0487704223707
Mn2  17.0062437680448   -11.7578538970877     2.3287001638882
Mn2  15.3062207895247    -8.8133273650489    12.0481039328923
O     6.0714020068665    -4.5404458428838     6.0928019870478
O     4.3713790283463    -1.5959193108450    15.8122057560519
O    11.1705230755416   -13.3723826474802     6.0921354975694
O     9.4705000970215   -10.4278561154414    15.8115392665735
O     3.1866245012928     2.9596263902512     6.0825499502425
O     1.4866015227727     5.9041529222900    15.8019537192466
O     8.2857455699680    -5.8723104143452     6.0818834607641
O     6.5857225914479    -2.9277838823064    15.8012872297682
O     0.9332430518142     1.6822662477821     6.0533422829054
O    -0.7667799267059     4.6267928099109    15.7727460519094
O     6.0323641204893    -7.1496705267243     6.0526757934269
O     4.3323411419692    -4.2051439946856    15.7720795624310
O     3.2237178578407    -2.8930326157643     6.1167021775932
O     1.5236948793206     0.0514939162745    15.8361059465972
O     8.3228389265159   -11.7249694203607     6.1160356881147
O     6.6228159479957    -8.7804428883219    15.8354394571188
O     0.9560828183334    -1.6400785528025     6.1059139787483
O    -0.7439401601868     1.3044479792362    15.8253177477523
O     6.0552038870085   -10.4720153573990     6.1052474892699
O     4.3551809084884    -7.5274888253602    15.8246512582739
O     2.4358377591511     1.6753737124261     3.6336974046790
O     0.7358147806310     4.6199002444649    13.3531011736831
O     7.5349588278263    -7.1565630921703     3.6330309152006
O     5.8349358493061    -4.2120365601315    13.3524346842046
O     5.2896851567785    -0.0275961155209     3.6537188209249
O     3.5896621782584     2.9169304165179    13.3731225899289
O    10.3888062254537    -8.8595329201173     3.6530523314465
O     8.6887832469335    -5.9150063880786    13.3724561004505
O     7.5526726325684    -4.5747704050830     3.6332762636462
O     5.8526496540482    -1.6302438730442    13.3526800326502
O    12.6517937012435   -13.4067072096794     3.6326097741678
O    10.9517707227234   -10.4621806776406    13.3520135431718
O     5.3055434074124    -5.8544180518367     3.6230239562230
O     3.6055204288923    -2.9098915197980    13.3424277252270
O    10.4046644760876   -14.6863548564332     3.6223574667445
O     8.7046414975674   -11.7418283243944    13.3417612357486
O     2.4608386110021    -1.6176108952995     3.6660149171243
```

| | | | |
|---|---|---|---|
| 0 | 0.7608156324819 | 1.3269156367393 | 13.3854186861283 |
| 0 | 7.5599596796772 | -10.4495476998959 | 3.6653484276458 |
| 0 | 5.8599367011571 | -7.5050211678571 | 13.3847521966499 |
| 0 | 7.5403325819054 | -1.3091258510045 | 3.6531329497328 |
| 0 | 5.8403096033852 | 1.6354006810343 | 13.3725367187368 |
| 0 | 12.6394536505805 | -10.1410626556009 | 3.6524664602544 |
| 0 | 10.9394306720604 | -7.1965361235621 | 13.3718702292584 |
| 0 | 2.4978512573595 | 4.2372562189895 | 8.4854761415347 |
| 0 | 0.7978282788394 | 7.1817827510283 | 18.2048799105387 |
| 0 | 7.5969723260347 | -4.5946805856069 | 8.4848096520563 |
| 0 | 5.8969493475145 | -1.6501540535681 | 18.2042134210603 |
| 0 | 7.4883025535042 | -1.5915339611152 | 8.2429693139533 |
| 0 | 5.7882795749841 | 1.3529925709236 | 17.9623730829573 |
| 0 | 12.5874236221794 | -10.4234707657116 | 8.2423028244749 |
| 0 | 10.8874006436592 | -7.4789442336728 | 17.9617065934789 |
| 0 | 0.0845810415709 | 2.9190086307584 | 8.3923959311201 |
| 0 | -1.6154419369492 | 5.8635351627972 | 18.1117997001242 |
| 0 | 5.1837021102461 | -5.9129281738380 | 8.3917294416417 |
| 0 | 3.4836791317259 | -2.9684016417992 | 18.1111332106458 |
| 0 | 5.0770705965302 | -0.2041833909277 | 8.2392533372218 |
| 0 | 3.3770476180100 | 2.7403431411111 | 17.9586571062258 |
| 0 | 10.1761916562053 | -9.0361201955241 | 8.2385868477434 |
| 0 | 8.4761686866852 | -6.0915936634853 | 17.9579906167474 |
| 0 | 2.4645954271232 | 1.5099660881686 | 8.3497027701300 |
| 0 | 0.7645724486031 | 4.4544926202074 | 18.0691065391341 |
| 0 | 7.5637164957984 | -7.3219707164278 | 8.3490362806516 |
| 0 | 5.8636935172782 | -4.3774441843890 | 18.0684400496556 |
| 0 | 2.4100055855722 | -1.6208004829549 | 8.5239843893594 |
| 0 | 0.7099826070520 | 1.3237260490839 | 18.2433881583634 |
| 0 | 7.5091266542473 | -10.4527372875513 | 8.5233178998810 |
| 0 | 5.8091036757272 | -7.5082107555125 | 18.2427216688850 |
| 0 | 8.3974045821384 | -5.8626317129236 | 1.3065747919767 |
| 0 | 6.6973825036183 | -2.9181051808848 | 11.0259785609807 |
| 0 | 13.4965265508135 | -14.6945685175200 | 1.3059083024983 |
| 0 | 11.7965035722934 | -11.7500419854812 | 11.0253120715023 |
| 0 | 0.9231650564783 | -1.5300643833802 | 1.3377417400967 |
| 0 | -0.7768579220418 | 1.4144621486586 | 11.0571455091007 |
| 0 | 6.0222861251535 | -10.3620011879766 | 1.3370752506183 |
| 0 | 4.3222631466334 | -7.4174746559378 | 11.0564790196223 |
| 0 | 6.0136079972739 | -7.2274847751323 | 1.2604905723121 |
| 0 | 4.3135850187537 | -4.2829582430936 | 10.9798943413161 |
| 0 | 11.1127290659490 | -16.0594215797287 | 1.2598240828336 |
| 0 | 9.4127060874289 | -13.1148950476900 | 10.9792278518377 |
| 0 | 6.0201102027811 | -4.4749332614079 | 1.3147290838105 |
| 0 | 4.3200872242609 | -1.5304067293691 | 11.0341328528145 |
| 0 | 11.1192312714562 | -13.3068700660043 | 1.3140625943321 |
| 0 | 9.4192082929361 | -10.3623435339655 | 11.0334663633361 |
| 0 | 6.0413900534684 | -1.3656500975223 | 1.2447879350410 |
| 0 | 4.3413670749483 | 1.5788764345165 | 10.9641917040451 |
| 0 | 11.1405111221435 | -10.1975869021187 | 1.2441214455626 |
| 0 | 9.4404881436234 | −7.2530603700799 | 10.9635252145666 |
| 0 | 3.3364854712871 | −2.8941070340477 | 1.3562565167706 |
| 0 | 1.6364624927670 | 0.0504194979910 | 11.0756602857746 |
| 0 | 8.4356065399622 | −11.7260438386442 | 1.3555900272922 |
| 0 | 6.7355835614421 | −8.7815173066054 | 11.0749937962962 |
| 0 | 11.1653017569011 | −1.6535923323214 | 6.0799170042454 |
| 0 | 9.4652787783809 | 1.2909341997174 | 15.7993207732494 |
| 0 | 16.2644228255762 | −10.4855291369178 | 6.0792505145670 |
| 0 | 14.5643998470561 | −7.5410026048790 | 15.7986542837710 |
| 0 | 11.1785918950263 | 1.6535706070225 | 6.0681128082813 |
| 0 | 9.4785689165062 | 4.5980971390613 | 15.7875165772854 |
| 0 | 16.2777129637015 | −7.1783661975739 | 6.0674463188029 |
| 0 | 14.5776899851813 | −4.2338396655351 | 15.7868500878069 |
| 0 | 8.3039769348279 | 5.8761691915909 | 6.0842019664315 |
| 0 | 6.6039539563077 | 8.8206957236296 | 15.8036057354355 |
| 0 | 13.4030980035030 | −2.9557676130056 | 6.0835354769531 |
| 0 | 11.7030750249829 | −0.0112410809668 | 15.8029392459571 |
| 0 | 6.0578791396479 | 4.5824549486078 | 6.0681855059117 |
| 0 | 4.3578561611277 | 7.5269814806466 | 15.7875892749157 |
| 0 | 11.1570002083230 | −4.2494818559886 | 6.0675190164332 |
| 0 | 9.4569772298029 | −1.3049553239498 | 15.7869227854373 |
| 0 | 8.2675466587014 | −0.0802857895686 | 6.1066749377331 |
| 0 | 6.5675236801813 | 2.8642407424702 | 15.8260787067371 |
| 0 | 13.3666677273766 | −8.9122225941650 | 6.1060084482547 |
| 0 | 11.6666447488564 | −5.9676960621262 | 15.8254122172587 |
| 0 | 5.9929750239030 | 1.2244454405915 | 6.1079580871536 |
| 0 | 4.2929520453828 | 4.1689719726302 | 15.8273618561576 |
| 0 | 11.0920960925781 | −7.6074913640049 | 6.1072915976752 |
| 0 | 9.3920731140580 | −4.6629648319662 | 15.8266953666792 |
| 0 | 7.5519530277913 | 4.5953937898702 | 3.6294260353689 |
| 0 | 5.8519300492712 | 7.5399203219089 | 13.3488298043730 |
| 0 | 12.6510740964664 | −4.2365430147262 | 3.6287595458905 |
| 0 | 10.9510511179463 | −1.2920164826875 | 13.3481633148946 |
| 0 | 10.4323459970390 | 2.9656467410863 | 3.6561766557800 |
| 0 | 8.7323230185188 | 5.9101732731251 | 13.3755804247840 |
| 0 | 15.5314670657141 | −5.8662900635101 | 3.6555101663016 |
| 0 | 13.8314440871940 | −2.9217635314714 | 13.3749139353056 |
| 0 | 12.6682816868520 | −1.6515564408778 | 3.6543492449422 |
| 0 | 10.9682587083318 | 1.2929700911609 | 13.3737530139462 |
| 0 | 17.7674027555271 | −10.4834932454743 | 3.6536827554637 |
| 0 | 16.0673797770070 | −7.5389667134355 | 13.3730865244678 |
| 0 | 10.4176613544500 | −2.9436822698560 | 3.6379080415031 |
| 0 | 8.7176383759299 | 0.0008442621828 | 13.3573118105072 |
| 0 | 15.5167824231252 | −11.7756190744524 | 3.6372415520247 |
| 0 | 13.8167594446051 | −8.8310925424137 | 13.3566453210287 |
| 0 | 7.5843740049639 | 1.3950857022530 | 3.5190610125200 |
| 0 | 5.8843510264438 | 4.3396122342917 | 13.2384647815240 |
| 0 | 12.6834950736390 | −7.4368511023435 | 3.5183945230415 |
| 0 | 10.9834720951189 | −4.4923245703047 | 13.2377982920456 |
| 0 | 2.4752704758464 | −4.2090762618233 | 3.6503451759262 |

```
0    0.7752474973263    -1.2645497297846    13.3697489449302
0    7.5743915445215   -13.0410130664198     3.6496786864477
0    5.8743685660014   -10.0964865343810    13.3690824554518
0    7.5837947352616     7.2445876717383     8.4288535332319
0    5.8837717567415    10.1891142037771    18.1482573022360
0   12.6829158039368    -1.5873491328581     8.4281870437535
0   10.9828928254166     1.3571773991807    18.1475908127576
0   12.6724353115694     1.5258854197144     8.4020284402986
0   10.9724123330493     4.4704119517532    18.1214322093027
0   17.7715563802446    -7.3060513848820     8.4013619508202
0   16.0715334017244    -4.3615248528432    18.1207657198242
0    5.1927542444039     5.8571950236458     8.3826450037341
0    3.4927312658838     8.8017215556845    18.1020487727382
0   10.2918753130791    -2.9747417809507     8.3819785142557
0    8.5918523345589    -0.0302152489119    18.1013822832597
0   10.2737378581223     2.8904264174695     8.3681021542043
0    8.5737148796022     5.8349529495083    18.0875059232084
0   15.3728589267975    -5.9415103871269     8.3674356647259
0   13.6728359482773    -2.9969838550881    18.0868394337299
0    7.5873961338659     4.4948959140044     8.3911311131904
0    5.8873731553458     7.4394224460432    18.1105348821944
0   12.6865172025411    -4.3370408905920     8.3904646237119
0   10.9864942240210    -1.3925143585532    18.1098683927160
0    7.5966502825796     1.3753995945015     8.6040212436857
0    5.8966273040595     4.3199261265403    18.3234250126897
0   12.6957713512548    -7.4565372100949     8.6033547542072
0   10.9957483727346    -4.5120106780561    18.3227585232113
0   13.5073001458697    -2.9198026454300     1.3248938236300
0   11.8072771673495     0.0247238866087    11.0442975926341
0   18.6064212145448   -11.7517394500264     1.3242273341516
0   16.9063982360247    -8.8072129179877    11.0436311031556
0    5.9825289795379     1.3927590028501     3.1235559589807
0    4.2825060010178     4.3372855348888    11.0317597279847
0   11.0816500482130    -7.4391778017463     1.3116894695022
0    9.3816270696929    -4.4946512697076    11.0310932385063
0   11.0951191579772    -4.2649354215963     1.2444216343746
0    9.3950961794570    -1.3204088895575    10.9638254033787
0   16.1942402266523   -13.0968722261927     1.2437551448962
0   14.4942172481322   -10.1523456941540    10.9631589139002
0   11.1386254764681    -1.5066935667847     1.3747381930345
0    9.4386024979479     1.4378329652540    11.0941419620386
0   16.2377465451432   -10.3386303713812     1.3740717035561
0   14.5377235666231    -7.3941038393424    11.0934754725602
0   11.1433439404342     1.6118042065468     1.3003128103053
0    9.4433209619140     4.5563307385856    11.0197165793094
0   16.2424650091093    -7.2201325980496     1.2996463208269
0   14.5424420305892    -4.2756060660109    11.0190500898310
0    8.4039367961331     0.0263615376035     1.3343557489030
0    6.7039138176129     2.9708880696422    11.0537595179070
0   13.5030578648082    -8.8055752669929     1.3336892594246
```

```
O  11.8030348862881   -5.8610487349542   11.0530930284286
```